\documentclass[onecolumn,numberedappendix]{emulateapj}
\usepackage[T1]{fontenc}
\usepackage{amsmath,times,amssymb,calligra}
\DeclareMathAlphabet{\mathcalligra}{T1}{calligra}{m}{n}
\DeclareFontShape{T1}{calligra}{m}{n}{<->s*[2.2]callig15}{}

\newcommand{\D}[2]{\frac{\partial #2}{\partial #1}}
\newcommand{\DD}[2]{\frac{\partial^2 #2}{\partial #1^2}}
\newcommand\bb[1]{\mbox{\boldmath{$#1$}}}
\newcommand\del{\bb{\nabla}}
\newcommand\bcdot{\,\bb{\cdot}\,}
\newcommand\btimes{\,\bb{\times}\,}
\newcommand{\mc}[1]{\mathcal{#1}}
\newcommand{\mcb}[1]{\bb{\mathcal{#1}}}

\newcommand{\msb}[1]{\bb{\mathsf{#1}}}
\newcommand{\eb}{\bb{b}}

\newcommand{\dr}[1]{\rho_{\rm {#1}}}
\newcommand{\nr}[1]{n_{\rm {#1}}}

\newcommand{\ngs}[2]{n_{{\rm g}^{#2}_{#1}}}
\newcommand{\inteng}{u_{\rm n}}

\newcommand{\vv}[1]{\bb{v}_{\rm {#1}}}
\newcommand{\vs}[1]{\bb{v}_{#1}}
\newcommand{\ww}[1]{\bb{w}_{\rm {#1}}}
\newcommand{\ws}[1]{\bb{w}_{#1}}

\newcommand{\tausn}[1]{\tau_{{#1}{\rm n}}}
\newcommand{\taurn}[1]{\tau_{\rm {#1}n}}
\newcommand{\tsinel}[1]{\tau_{{#1}{\rm ,inel}}}
\newcommand{\trinel}[1]{\tau_{\rm {#1},inel}}

\newcommand{\chif}{\chi_{\mc{F}}}
\newcommand{\chir}{\chi_{\rm R}}
\newcommand{\kappap}{\kappa_{\rm P}}
\newcommand{\kappae}{\kappa_{\rm E}}

\newcommand{\denomone}{1-\sum_k\frac{\tau_0}{\tsinel{k}}\frac{\varrho_k}{1+\varrho_k}}
\newcommand{\denomtwo}{\frac{\tau_0}{\taurn{g_0}} + \frac{\tau_0}{\trinel{g_+}}\frac{1}{1+\varrho_{\rm g_+}} + \frac{\tau_0}{\trinel{g_-}}\frac{1}{1+\varrho_{\rm g_-}}}
\newcommand{\denomthree}{1-\frac{\omega^2_{\rm g_+}\tausn{g_+}^2\vartheta_{\rm g_- g_+}}{1+\omega^2_{\rm g_+}\tausn{g_+}^2(1-\varphi_{\rm g_+})}\frac{\omega^2_{\rm g_-}\tausn{g_-}^2\vartheta_{\rm g_+ g_-}}{1+\omega^2_{\rm g_-}\tausn{g_-}^2(1-\varphi_{\rm g_-})}}

\begin{document}
\title[Nonisothermal Magnetic Star Formation]{The Nonisothermal Stage of Magnetic Star Formation. \\
    I. Formulation of the Problem and Method of Solution}
\author{Matthew W. Kunz and Telemachos Ch. Mouschovias}
\affil{Departments of Physics and Astronomy, University of
Illinois at Urbana-Champaign,
    1002 W. Green Street, Urbana, IL 61801}

\begin{abstract}
We formulate the problem of the formation and subsequent evolution
of fragments (or cores) in magnetically-supported,
self-gravitating molecular clouds in two spatial dimensions. The
six-fluid (neutrals, electrons, molecular and atomic ions,
positively-charged, negatively-charged, and neutral grains)
physical system is governed by the radiative, nonideal
magnetohydrodynamic (RMHD) equations. The magnetic flux is not
assumed to be frozen in any of the charged species. Its evolution
is determined by a newly-derived generalized Ohm's law, which
accounts for the contributions of both elastic and inelastic
collisions to ambipolar diffusion and Ohmic dissipation. The
species abundances are calculated using an extensive
chemical-equilibrium network. Both MRN and uniform grain size
distributions are considered. The thermal evolution of the
protostellar core and its effect on the dynamics are followed by
employing the grey flux-limited diffusion approximation. Realistic
temperature-dependent grain opacities are used that account for a
variety of grain compositions. We have augmented the
publicly-available Zeus-MP code to take into consideration all
these effects and have modified several of its algorithms to
improve convergence, accuracy and efficiency. Results of magnetic
star formation simulations that accurately track the evolution of
a protostellar fragment from a density $\simeq 10^{3}$ cm$^{-3}$
to a density $\simeq 10^{15}$ cm$^{-3}$, while rigorously
accounting for both nonideal MHD processes and radiative transfer,
are presented in a separate paper.
\end{abstract}


\section{Introduction --- Background}\label{section:intro}

The formulation of a theory of star formation is a formidable
task. It requires understanding of the nonlinear interactions
among self-gravity, magnetic fields, rotation, chemistry
(including grain effects), turbulence, and radiation. Stars form
in fragments within interstellar molecular clouds, which have
sizes ranging from 1 to 5 pc, masses from a few tens to $10^5$
M$_\odot$, mean densities $\simeq 10^3$ cm$^{-3}$, and
temperatures $\simeq 10$ K \citep{myers85,heiles87}. Their
spectral lines have Doppler-broadened linewidths that suggest
supersonic (but subAlfv\'{e}nic) internal motions. In the deep
interiors of such clouds, high-energy cosmic rays ($>100$ MeV)
maintain a degree of ionization $x_{\rm i} \lesssim 10^{-7}$,
whereas ultraviolet (UV) ionization is responsible for a much
greater degree of ionization $x_{\rm i}\gtrsim 10^{-5}$ in the
outer envelopes.

\subsection{Magnetic Fields and Star Formation}

The possible importance of magnetic fields to the support of
interstellar clouds and to the regulation of star formation was
first studied by \citet{cf53}, \citet{ms56}, and \citet{mestel65}
using the virial theorem. Similar investigations by
\citet{strittmatter66a,strittmatter66b} and \citet{spitzer68}
followed. \citet{mestel66} calculated the magnetic forces on a
spherically-symmetric, gravitationally-bound cloud.
Self-consistent calculations by
\citet{mouschovias76a,mouschovias76b} produced exact equilibria of
initially uniform, isothermal, magnetic clouds embedded in a hot
and tenuous, electrically-conducting external medium. \citet{ms76}
used these equilibrium states to find the critical mass-to-flux
ratio  $(M/\Phi_{\rm B})_{\rm cr} = 1/(63G)^{1/2}$ that must be
exceeded for collapse against the magnetic forces to set in.
\citet{sb80} performed numerical simulations of the collapse of a
supercritical (as a whole) magnetic cloud. A picture of molecular
clouds emerged in which magnetic fields play a central role in
their support and evolution \citep{mouschovias78}. To this date,
magnetic braking remains the only mechanism that has been shown
quantitatively to resolve the most significant dynamical problem
of star formation, namely, the angular momentum problem
(Mouschovias \& Paleologou 1979, 1980; see also summary below).

Subsequent observations lent credence to this picture by revealing
the importance of magnetic fields through both dust polarization
measurements and Zeeman observations. Polarization studies have
exhibited large-scale ordered magnetic fields connecting
protostellar cores to their surrounding envelopes
\citep{vct81,hvsssgs87,nghpd89,nddgh97,lcgr01,lgc03,cnwtk04,afg08},
often with an hourglass morphology
\citep{schleuning98,hddsv99,gcr99,svhdndd00,lcgr02,cc06,grm06}, as
predicted by the theoretical calculations \citep{mouschovias76b,fm93}.
A large body of Zeeman observations
\citep{ck83,kc86,tck86,ctk87,gchmt89,ctghkm93,cmtc94,crmt96,tcghkm96,
crtg99,ctlpk99,crutcher99,hc05,ccw05}
revealed magnetic fields in the range $\simeq \, 10 - 200 \, \mu$G
in molecular clouds, from small isolated ones to massive
star-forming ones. These values are more than sufficient to
establish the importance of magnetic fields in molecular cloud
dynamics.

It was recognized early on (e.g., see Babcock \& Cowling 1953, p.
373) that the magnetic flux of an interstellar blob of mass
comparable to a stellar mass is typically several orders of
magnitude greater than that of magnetic young stars. This is the
so-called ``magnetic flux problem" of star formation. It lies in
the fact that substantial flux loss must take place at some stage
during star formation. Ambipolar diffusion (the relative motion
between plasma and neutrals) was first proposed by \citet{ms56} as
a means by which an interstellar cloud as a whole would reduce its
magnetic flux and thereby collapse. \citet{pm65}
undertook a detailed calculation of the collapse of such
(spherical) cloud. \citet{spitzer68} calculated the
ambipolar-diffusion timescale by assuming that the magnetic force
on the ions is balanced by the (self-)gravitational force on the
neutrals. \citet{nakano79} followed the quasistatic contraction of
a cloud due to ambipolar diffusion using a sequence of
Mouschovias' (1976b) equilibrium states, each one of which had a
smaller magnetic flux than the previous one.

A new solution for ambipolar diffusion by \citet{mouschovias79}
showed that the essence of this process is a redistribution of
mass in the central flux tubes of a molecular cloud, rather than a
loss of magnetic flux by the cloud as a whole. He found the
ambipolar-diffusion timescale to be typically three orders of
magnitude smaller in the interior of a cloud than in the outermost
envelope, where there is a much better coupling between neutral
particles and the magnetic field because of the much greater
degree of ionization. This suggested naturally a self-initiated
fragmentation of (or core formation in) molecular clouds on the
ambipolar-diffusion timescale
\begin{equation*}
\tau_{\rm AD}=1.8\times 10^6\,\left(\frac{x_{\rm
i}}{10^{-7}}\right)\;{\rm yr}\,
\end{equation*}
(Mouschovias 1979, eq. [37]). The inefficiency of star formation
was thereby attributed to the self-initiated formation and
contraction of molecular cloud fragments (or cores) due to
ambipolar diffusion in otherwise magnetically supported clouds
\citep{mouschovias76b,mouschovias77,mouschovias78,mouschovias79}.
The {\em central} mass-to-flux ratio eventually exceeds its
critical value for collapse,
\begin{equation*}
\left(\frac{dm}{d\Phi_{\rm B}}\right)_{\rm c,cr} =
\frac{3}{2}\left(\frac{M}{\Phi_{\rm B}}\right)_{\rm cr}\,,
\end{equation*}
(see Mouschovias 1976a, eq. [44]), and dynamic contraction ensues.
Detailed numerical calculations in slab \citep{pm83,mpf85},
cylindrical \citep{mm91,mm92a,mm92b}, and axisymmetric geometries
\citep{fm92,fm93,cm93,cm94,cm95,bm94,bm95a,bm95b,ck98,dm01,tm05a,tm05b,tm07a,tm07b,tm07c}
transformed this scenario of star formation into a theory with
predictive power.

\subsection{Rotation}

During the early, isothermal phase of star formation, a cloud (or
a fragment) must also lose a large fraction of its angular
momentum (e.g., see Spitzer 1968, p. 231). Observations show that
molecular clouds and embedded fragments (or cores) rarely exhibit
rotation significantly greater than that of the background medium
\citep{ga85}. If angular momentum were conserved from the initial
galactic rotation (i.e., starting from angular velocity
$\Omega_0\simeq 10^{-15}$ s$^{-1}$ at the mean density of the
interstellar medium $\simeq 1$ cm$^{-3}$), centrifugal forces
would not allow even the formation of interstellar clouds
(Mouschovias 1991, \S~2). Fragmentation does not alter this
conclusion (Mouschovias 1977, \S~1). This is referred to as the
``angular momentum problem" of star formation. As far as clouds
and their cores are concerned, the angular momentum problem has
been shown to be resolved by magnetic braking (i.e., the transport
of angular momentum from a fragment to its surrounding medium
through the propagation of torsional Alfv\'{e}n waves along
magnetic field lines connecting the fragment to the cloud
envelope) analytically by \citet{mp79,mp80} and numerically by
\citet{bm94,bm95a,bm95b} and \citet{ml08}. Hence the centrifugal forces resulting
from the cloud's or core's rotation have a negligible effect on
the evolution of the contracting core, at least up to central
densities of $\approx 10^{14}$ cm$^{-3}$ (see the last paragraph
of Tassis \& Mouschovias 2007b).

\subsection{Grain Effects}

Interstellar grains comprise about 1\% of the mass in the
interstellar medium \citep{spitzer78}. \citet{baker79} and
\citet{elmegreen79} suggested that charged grains may couple to
the magnetic field and thereby play a role in ambipolar diffusion
and star formation. \citet{elmegreen79} and \citet{nu80} compared
and ambipolar-diffusion timescale and the free-fall timescale and
concluded that ambipolar diffusion occurs over too long a
timescale (roughly 10 times greater than free-fall) to be a
relevant process in star formation. Refinements by the same
authors \citep{elmegreen86,un90,nnu91} led to similar conclusions.
Through detailed numerical simulations of core formation and
evolution including the effects of (negative and neutral) dust
grains, \citet{cm93,cm94} found that grains lengthen the timescale
for the formation of a core because of grain-neutral collisions,
but cautioned that the ambipolar-diffusion timescale should not be
compared to the free-fall timescale in determining its relevance
in magnetically-supported clouds, as originally pointed out by
\citet{mouschovias77}, because molecular clouds are not free-falling.
Velocities characteristic of such collapse have not been observed.
\citet{cm95} extended these calculations by including UV
ionization and a variety of atomic metal ions (C$^+$, S$^+$,
Si$^+$, Mg$^+$, Na$^+$, Fe$^+$). Attention was also paid to the
complementary effect of protostellar evolution on the microscopic
physics and chemistry \citep{cm96,cm98}.

\subsection{MHD Waves and/or Turbulence}

The extent to which turbulence (or waves) may or may not affect
the evolution of a protostellar fragment has been a topic of
debate for several decades, receiving increased attention in
recent years \citep{mlk04,mtk06}. A consensus has yet to be
reached concerning even what causes and maintains the observed
broad linewidths long thought to be indicative of supersonic
turbulence
\citep{ze74,am75,larson81,zj83,mouschovias87,mp95,mg99,es04,mtk06},
although \citet{mp95} and \citet{mtk06} showed quantitatively that
observations contradict the key
assumption of turbulence simulations, namely, that molecular
clouds are magnetically supercritical by a factor $4 - 10$.
Despite a lack of agreement on the origin of the linewidths,
analytical \citep{mouschovias91} and numerical calculations (Eng
2002; Eng \& Mouschovias 2009, in preparation) have demonstrated
that turbulence (or waves) plays an insignificant role in the star
formation process once dynamical contraction of a fragment (or
core) ensues. Observations showing narrowing and eventual
thermalization of linewidths in protostellar cores
\citep{bpddc81,mb83,mlb83,bapabwt00} are in agreement with this
conclusion and with an earlier version of it \citep{mouschovias87}.

\subsection{Radiative Transfer}

During the early phases of star formation the energy produced by
compressional heating is radiated away by the dust grains in the
infrared. At higher densities ($\simeq 10^{11}$ cm$^{-3}$), the
core traps and retains part of this heat and its temperature
begins to rise. The evolution of the temperature in this
nonisothermal regime may be approximated (but substantially
overestimated) by using an adiabatic equation of state
\citep{boss81,dm01,tm07a,tm07b,tm07c}. More realistic equations of
state have also been employed by, for example, \citet{bate98}. To
accurately model the nonisothermal phase of protostellar
contraction, however, one needs to include a proper treatment of
radiative transfer. Early efforts to include radiative transfer in
(nonmagnetic) star formation calculations were confined to the use
of the diffusion approximation
\citep{bodenheimer68,larson69,larson72,bb75,bb76,tscharnuter75,yk77}.
While the diffusion approximation is strictly applicable only to
optically thick regions, its ease of implementation and relatively
low computational cost make it an attractive choice. The Eddington
approximation offers a slight improvement in that it retains some
of the rigor of using moments of the radiative transfer equation,
while making the simplifying assumption that the radiation field
is everywhere isotropic. Its use in numerical calculations of
(nonmagnetic) star formation has been documented in
\citet{tscharnuter78}, \citet{tw79}, \citet{wn80a,wn80b},
\citet{boss84,boss86,boss88}, and \citet{bm95}.
By implicitly assuming that photons always travel a
distance comparable to their mean-free path (even if this distance
exceeds the free-flight distance $c\Delta t$, where $\Delta t$ is
the computational timestep), the Eddington approximation gives
unphysical behavior in optically thin regions, in which the
mean-free-path is huge. The result is a signal speed unbound by
the speed of light, i.e., it violates causality (e.g., see \S~97
of Mihalas \& Mihalas 1984).

Increasing the accuracy and realism of a radiative transfer algorithm
often requires making limiting assumptions about the hydrodynamics in
order to make the problem tractable (e.g., Yorke 1980; Masunaga et al.
1998). A full frequency- and angle-dependent treatment of the radiation
is nearly always confined to postprocessing the results of a hydrodynamic
calculation \citep{yorke77,ys81,as85,as86} or a grey (i.e., independent
of frequency) radiation hydrodynamic calculation \citep{by90,byrt90}. By
contrast, the flux-limited diffusion (FLD) approximation \citep{lp81} is
a propitious compromise that retains some of the advantages of the
diffusion and Eddington approximations, while preserving causality and
coupling self-consistently to the hydrodynamic equations.

\subsection{Outline}

In this paper we formulate the problem of the formation and
evolution of protostellar fragments (or cores) in
magnetically-supported, self-gravitating molecular clouds,
including the effects of both ambipolar diffusion and Ohmic
dissipation (which becomes important at high densities), grain
chemistry and dynamics, and radiation. Using the results of Eng
(2002) and Eng \& Mouschovias (2009, in preparation), and
\citet{bm94}, we may safely ignore the effects of turbulence and
rotation, respectively, on the evolution of the protostellar core
for the densities considered here. The physical and chemical
properties of the model cloud are summarized in Section
\ref{section:modelcloud}. The radiation magnetohydrodynamic (RMHD)
equations governing the evolution of the model cloud are presented
and discussed in Section \ref{section:equations}. In Section
\ref{section:chem} we present the chemical model used in the
calculations. The physics of magnetic diffusion (ambipolar and
ohmic) is handled by using a generalized Ohm's law, which is
derived in Section \ref{section:fluxloss}. We treat the radiative
transfer using the grey FLD approximation, with realistic grain
opacities accounting for a variety of grain compositions
(\S~\ref{section:radtrans}). The numerical method of solution is
discussed in Section \ref{section:zeusmp}. Finally, we give the
simplified set of equations and a brief summary in Section
\ref{section:summary}. Details, mostly mathematical, are left for
the Appendix. Results are presented in a separate paper
(Kunz \& Mouschovias 2009, in preparation).

\section{Basic Properties of the Model Cloud}\label{section:modelcloud}

\begin{table}
\centering \caption{\label{table:chem} Chemical reaction network
used in the calculation of the abundances of charged species.}
\begin{minipage}{\textwidth}
\centering
\begin{tabular}{l c c c}
\hline
\multicolumn{4}{c}{Relevant Chemical Reactions in Molecular Clouds}\\
\hline
Cosmic-Ray Ionization:            & ${\rm H_2 + CR}$ & $\rightarrow$ & ${\rm H_2^+ + e}$ \\
                                  & ${\rm H_2^+ +  H_2}$ & $\rightarrow$ & ${\rm H_3^+ +  H} $\\
                                  & ${\rm H_3^+ + CO}$ & $\rightarrow$ & ${\rm HCO^+ + H_2}$ \\
Dissociative Recombination:       & ${\rm HCO^{+} + e}$ & $\rightarrow$ &${\rm H + CO} $\\
Radiative recombination: \footnote{\centering \;A$^{+}$ represents an atomic
ion, such as Na$^{+}$, Mg$^{+}$, and K$^{+}$, and A$^{0}$ the
corresponding neutral atom.}
                                  & ${\rm A^{+} + e}$ & $\rightarrow$ & ${\rm A^{0}} + h{\rm \nu}$ \\
Charge transfer: \footnotemark[$a$]
                                  & ${\rm A^0 + HCO^{+}}$ & $\rightarrow$ &${\rm A^{+} + HCO}$ \\
$e^-$ attachment onto grains:     & ${\rm e + g_{0}}$ & $\rightarrow$ & ${\rm g_{-}} $  \\
                                  & ${\rm e + g_{+}}$ & $\rightarrow$ & ${\rm g_{0}} $  \\
Atomic-ion attachment onto grains: \footnotemark[$a$]
                                  & ${\rm A^{+} + g_{-}}$ & $\rightarrow$ & ${\rm A^0 + g_{0}} $   \\
                                  & ${\rm A^{+} + g_{0}}$ & $\rightarrow$ & ${\rm A^0 + g_{+}}$ \\
Molecular-ion attachment onto grains: & ${\rm HCO^{+} + g_{0}}$ & $\rightarrow$ & ${\rm HCO + g_{+}}$ \\
                                  & ${\rm HCO^{+} + g_{-}}$ & $\rightarrow$ & ${\rm HCO + g_{0}}$  \\
Charge transfer by grain-grain collisions:
                                  & ${\rm g}^\alpha_+ + {\rm g}^{\alpha'}_-$ & $\rightarrow$ & ${\rm g}^\alpha_0 + {\rm g}^{\alpha'}_0$ \\
                                  & ${\rm g}^{\alpha}_\pm + {\rm g}^{\alpha'}_0$ & $\rightarrow$ & ${\rm g}^\alpha_0 + {\rm g}^{\alpha'}_\pm$\\
\hline
\end{tabular}
\end{minipage}
\end{table}

We consider a self-gravitating, magnetic, weakly-ionized model
molecular cloud consisting of neutral particles (H$_2$ with 20\%
He by number), ions (both molecular HCO$^+$ and atomic Na$^+$,
Mg$^+$, K$^+$), electrons, singly negatively-charged grains,
singly positively-charged grains, and neutral grains. Following
\citet{dm01}, the abundances of all species (except the neutrals)
are determined from the chemical reaction network shown in Table
\ref{table:chem} and described below in Section
\ref{subsection:chem}. Cosmic rays of energy $\gtrsim$ 100 MeV are
mainly responsible for the degree of ionization in the cloud. Once
column densities $\gtrsim$ 100 g cm$^{-2}$ are achieved, cosmic
rays are appreciably attenuated. At even higher densities, cosmic
rays are effectively shielded and radioactive decays become the
dominant source of ionization. Finally, at temperatures on the
order of $1000\,{\rm K}$ or higher, thermal ionization of
potassium becomes important. UV radiation provides an additional
ionization mechanism, but it only affects the outer envelope of
molecular clouds \citep{hws71,gl74}. We consider spherical grains
whose radii are determined by either a uniform or an MRN
\citep{mrn77} size distribution. In the case of collisions of ions
(molecular or atomic) with grains, we assume that the ions do not
get attached to the grains, but rather that they get neutralized,
with the resulting neutral particle escaping into the gas phase.
Thus the total abundance of metals as well as the total HCO
abundance remain constant. Grain growth is not considered here.

We follow the ambipolar-diffusion--initiated evolution of an
axisymmetric two-dimensional model cloud from typical mean
molecular cloud densities ($\simeq 300$ cm$^{-3}$) to densities
characteristic of the formation of a hydrostatic protostellar
core. The axis of symmetry is aligned with the $z$-axis of a
cylindrical polar coordinate system ($r$, $\phi$, $z$).
Isothermality is an excellent approximation for the early stages
of star formation, while the density is smaller than $\approx
10^{10}$ cm$^{-3}$ \citep{gaustad63,hayashi66,larson69}. However,
once the heat generated by released gravitational energy during
core collapse is unable to escape freely (at a central number
density of $n_{\rm opq} \simeq 10^7$ cm$^{-3}$), radiative
transfer calculations are employed to determine the thermal
evolution of the core.\footnote{We have varied the density $n_{\rm
opq}$ at which we turn on the radiative transfer solver from
$10^6-10^{11}$ cm$^{-3}$ and found that $n_{\rm opq}\lesssim 10^7$
cm$^{-3}$ is necessary to achieve a smooth transition from
isothermality. This numerical necessity does {\it not} mean that
the isothermality assumption breaks down at as low a density as
$10^7$ cm$^{-3}$.} This is an improvement over previous magnetic
star formation calculations to reach these densities
\citep{dm01,tm07a,tm07b,tm07c}, which assumed an adiabatic
equation of state beyond a critical density because of the high
computational expense of radiative transfer calculations.
Numerical techniques and computer hardware have matured enough by
now to render these once impractical calculations feasible.

\section{The Six-Fluid RMHD Description of Magnetic Star Formation}\label{section:equations}

The RMHD equations governing the behavior of the six-fluid system
(neutrals, electrons, ions, negative, positive, and neutral
grains) are:
\begin{subequations}
\begin{equation}\label{eqn:nzcont}
\D{t}{\dr{n}} + \del\bcdot(\dr{n}\vv{n}) = 0\,,
\end{equation}
\begin{equation}\label{eqn:gtcont}
\D{t}{(\dr{g_-}+\dr{g_0}+\dr{g_+})} +
\del\bcdot(\dr{g_-}\vv{g_-}+\dr{g_0}\vv{g_0}+\dr{g_+}\vv{g_+}) =
0\,,
\end{equation}
\begin{equation}\label{eqn:nzforce}
\D{t}{(\dr{n}\vv{n})} + \del\bcdot(\dr{n}\vv{n}\vv{n}) = -\del
P_{\rm n} -\rho\del\psi + \frac{1}{c}\bb{j}\btimes\bb{B} +
\frac{1}{c}\chif\mcb{F}\,,
\end{equation}
\begin{equation}\label{eqn:emforce}
0 = -e\nr{e}\left(\bb{E}+\frac{\vv{e}}{c}\btimes\bb{B}\right) +
\bb{F}_{\rm en}\,,
\end{equation}
\begin{equation}\label{eqn:ipforce}
0 = +e\nr{i}\left(\bb{E}+\frac{\vv{i}}{c}\btimes\bb{B}\right) +
\bb{F}_{\rm in}\,,
\end{equation}
\begin{equation}\label{eqn:gmforce}
0 = -e\nr{g_-}\left(\bb{E}+\frac{\vv{g_-}}{c}\btimes\bb{B}\right)
+ \bb{F}_{\rm g_-n} + \bb{F}_{\rm g_-g_0,inel}\,,
\end{equation}
\begin{equation}\label{eqn:gpforce}
0 = +e\nr{g_+}\left(\bb{E}+\frac{\vv{g_+}}{c}\btimes\bb{B}\right)
+ \bb{F}_{\rm g_+n} + \bb{F}_{\rm g_+g_0,inel}\,,
\end{equation}
\begin{equation}\label{eqn:gzforce}
0 = \bb{F}_{\rm g_0n} + \bb{F}_{\rm g_0g_-,inel} + \bb{F}_{\rm
g_0g_+,inel}\,,
\end{equation}
\begin{equation}\label{eqn:ampere}
\del\btimes\bb{B} = \frac{4\pi}{c}\bb{j}\,,
\end{equation}
\begin{equation}\label{eqn:current}
\bb{j} =
e\left(\nr{i}\vv{i}-\nr{e}\vv{e}+\nr{g_+}\vv{g_+}-\nr{g_-}\vv{g_-}\right)\,,
\end{equation}
\begin{equation}\label{eqn:faraday}
\D{t}{\bb{B}} = -c\del\btimes\bb{E}\,,
\end{equation}
\begin{equation}\label{eqn:poisson}
\del^2\psi = 4\pi G\dr{n}\,,
\end{equation}
\begin{equation}\label{eqn:egcont}
\D{t}{\inteng} + \del\bcdot(\inteng\vv{n}) = -P_{\rm
n}\del\bcdot\vv{n} - 4\pi\kappap\mc{B} + c\kappae\mc{E} +
\Gamma_{\rm diff}\,,
\end{equation}
\begin{equation}\label{eqn:ercont}
\D{t}{\mc{E}} + \del\bcdot(\mc{E}\vv{n}) = -\del\bcdot\mcb{F} -
\del\vv{n}\,\bb{:}\,\msb{P} + 4\pi\kappap\mc{B} -
c\kappae\mc{E}\,,
\end{equation}
\begin{equation}\label{eqn:fluxcont}
\D{t}{\mcb{F}} + \del\bcdot(\mcb{F}\vv{n}) = -c^2\del\bcdot\msb{P}
-c\chif\mcb{F}\,.
\end{equation}
\end{subequations}
The quantities $\rho_s$, $n_s$, and $\vs{s}$ refer to the mass
density, number density, and velocity of species $s$; the
subscripts n, i, e, g$_-$, g$_+$, and g$_0$ refer, respectively,
to the neutrals, ions, electrons, negatively-charged grains,
positively-charged grains, and neutral grains. The quantities
$\bb{E}$ and $\bb{B}$ denote the electric and magnetic field,
respectively, $\bb{j}$ the total electric current density,
$\inteng$ the internal energy density, $P_{\rm n}$ the gas
pressure, and $\psi$ the gravitational potential. The source term
$\Gamma_{\rm diff}$ in the internal energy equation
(\ref{eqn:egcont}) represents heating due to ambipolar diffusion
and Ohmic dissipation (see \S~\ref{subsection:joule} below). The
magnetic field satisfies the condition $\del\bcdot\bb{B}=0$
everywhere at all times. The definitions and derivations of
$\bb{F}_{s{\rm n}}$ (the frictional force per unit volume on
species $s$ due to elastic collisions with neutrals) and
$\bb{F}_{\gamma\delta{\rm ,inel}}$ (the force per unit volume on
grain fluid $\gamma$ due to the conversion of dust particles of
fluid $\delta$ into dust particles of fluid $\gamma$) are
discussed in detail in \S~3 of \citet{tm05a}, as well as in
\citet{cm93} and \citet{mouschovias96} (in the absence of
positively-charged grains).

The radiation variables are the Planck function $\mc{B}$, the
total (frequency-integrated) radiation energy density $\mc{E}$,
the total (frequency-integrated) radiation momentum density
$\mc{F}$, and the total (frequency-integrated) radiation pressure
tensor $\msb{P}$:
\begin{subequations}
\begin{equation}
\mc{E}(\bb{x},t) = \frac{1}{c}\int_0^\infty d\nu \oint d\Omega\,
I(\bb{x},t;\bb{\Omega},\nu)\,,
\end{equation}
\begin{equation}
\mcb{F}(\bb{x},t) = \int_0^\infty d\nu \oint d\Omega\,
I(\bb{x},t;\bb{\Omega},\nu)\, \hat{\bb{n}} \,,
\end{equation}
\begin{equation}
\msb{P}(\bb{x},t) = \frac{1}{c}\int_0^\infty d\nu \oint d\Omega\,
I(\bb{x},t;\bb{\Omega},\nu)\, \hat{\bb{n}} \hat{\bb{n}} \,.
\end{equation}
\end{subequations}
Here we have introduced the frequency $\nu$, the extinction
coefficient (i.e., opacity) $\chi(\nu)$ [$\equiv\kappa(\nu) +
\sigma(\nu)$, where $\kappa$ is the absorption coefficient and
$\sigma$ the scattering coefficient], and the radiation specific
intensity $I$. The material properties $\kappap$, $\kappae$, and
$\chif$ are the Planck and energy mean absorption coefficients,
and the flux-weighted mean opacity, respectively; they are given by
\begin{equation}
\renewcommand{\theequation}{\arabic{equation}a,b,c}
\kappap \equiv
\frac{1}{\mc{B}}\int_0^\infty\kappa(\nu)\mc{B}(\nu)d\nu\,,\qquad
\kappae \equiv
\frac{1}{\mc{E}}\int_0^\infty\kappa(\nu)\mc{E}(\nu)d\nu\,,\qquad
\chif   \equiv
\frac{1}{\mcb{F}}\int_0^\infty\chi(\nu)\mcb{F}(\nu)d\nu\,.
\end{equation}
Equations (\ref{eqn:ercont}) and (\ref{eqn:fluxcont}) are obtained
from taking moments of the radiation transport equation
\begin{equation}\label{eqn:radtrans}
\left(\frac{1}{c}\D{t}{} + \bb{\Omega}\bcdot\del\right)
I(\bb{x},t;\bb{\Omega},\nu) =
\chi(\bb{x},t;\bb{\Omega},\nu)\bigl[S(\bb{x},t;
\bb{\Omega},\nu)-I(\bb{x},t;\bb{\Omega},\nu)\bigr]
\end{equation}
under the assumptions that all the radiation variables are
measured in the comoving frame of the fluid (in this frame the
material properties are isotropic) and that the material
properties are grey \citep{mm84}.
\footnote{We caution here that \citet{psy95} and \citet{ys02} have
shown that multi-frequency calculations generally produce higher
dust temperatures and greater degrees of anisotropy in the radiation
field than corresponding grey calculations.}
We have taken the source function $S$ in the transport equation
(\ref{eqn:radtrans}) to be given by
\begin{equation}
4\pi S_\nu = \frac{4\pi\kappa_\nu\mc{B}_\nu +
c\sigma_\nu\mc{E}_\nu}{\kappa_\nu+\sigma_\nu}\,,
\end{equation}
taking into account both establishment of local thermodynamic
equilibrium and coherent isotropic scattering of radiation
\citep{mm84}.

In the force equations for the electrons, ions, and grains, the
acceleration terms have been neglected due to the small inertia of
these species. The acceleration term for the plasma was included
by \citet{mpf85} and it was shown that the plasma reaches a
terminal drift velocity very fast. Similarly, the thermal-pressure
and gravitational forces have been dropped from the force
equations of all species other than the neutrals because they are
negligible compared to the electromagnetic and collisional forces.
The inelastic momentum transfer by the electron and ion fluids due
to attachment onto grains and neutralization are negligible
compared to the momentum transfer due to elastic collisions, and
they have been omitted from the force equations
(\ref{eqn:emforce}) and (\ref{eqn:ipforce}) (see discussion in
\S~3.1 of Ciolek \& Mouschovias 1993). As we
consider a distribution of grain sizes, it should be noted that
equations (\ref{eqn:gtcont}), (\ref{eqn:gmforce}) -
(\ref{eqn:gzforce}) apply to each grain size separately.

\begin{figure}
\epsscale{0.4} \plotone{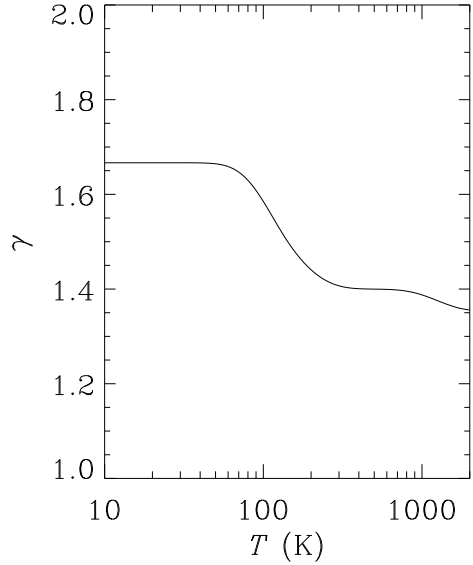} \caption{Dependence of
$\gamma$ (ratio of specific heats) on temperature for
H$_2$.}\label{fig:gamma}
\end{figure}

The full set of RMHD equations are closed with constitutive
relations for the gas pressure, opacities, and the Planck function
[i.e., $P_{\rm n}=P_{\rm n}(\dr{n},T)$, $\chif=\chif(\dr{g},T)$,
$\kappae=\kappae(\dr{g},T)$, $\kappap=\kappap(\dr{g},T)$, and
$\mc{B}=\mc{B}(T)$, where $T$ is the gas temperature and
$\dr{g}\equiv\dr{g_-}+\dr{g_0}+\dr{g_+}$ is the total grain mass
density]. In addition, we close the radiation moment equations
with the tensor variable Eddington factor $\msb{f}$ which is used
to eliminate the radiation stress tensor $\msb{P}$ in favor of the
radiation energy density $\mc{E}$ via
\begin{equation}\label{eqn:ptensor}
\msb{P} = \msb{f}\mc{E}\,.
\end{equation}
The Eddington factor $\msb{f}$ is determined by employing the
flux-limited diffusion approximation (see \S~\ref{subsection:FLD}
below). The equation of state for an ideal gas is given by
\begin{equation}\label{eqn:eos}
P_{\rm n} = (\gamma-1)\inteng\,,
\end{equation}
where $\gamma = k_{\rm B}/c_{\rm V} + 1$ is the adiabatic index
and $c_{\rm V}$ is the specific heat at constant volume per
particle:
\begin{equation}
c_{\rm V} = \frac{3}{2}k_{\rm B} + c^{\rm vib}_{\rm V} + c^{\rm
rot}_{\rm V} \,,
\end{equation}
assuming that there is no coupling between the rotational and
vibrational degrees of freedom of the molecule (in this case,
H$_2$). The vibrational specific heat is
\begin{equation}
c^{\rm vib}_{\rm V} = k_{\rm B}\left(\frac{\Theta_{\rm
vib}}{T}\right)^2\frac{\exp(\Theta_{\rm
vib}/T)}{\bigl[\exp(\Theta_{\rm vib}/T)-1\bigr]^2}\,,
\end{equation}
where $\Theta_{\rm vib}=6100\,{\rm K}$; the rotational specific
heat for a 3:1 mixture of ortho- and para-hydrogen is
\begin{equation}
c^{\rm rot}_{\rm V} = \frac{3}{4}k_{\rm B}x^2\DD{x}{}\ln Z_{\rm 0}
+ \frac{1}{4}k_{\rm B}x^2\DD{x}{}\ln Z_{\rm p}\,,
\end{equation}
where $Z_{\rm o}$ and $Z_{\rm p}$ are the ortho- and para-hydrogen
partition functions, respectively, given by
\begin{equation}
\renewcommand{\theequation}{\arabic{equation}a,b}
Z_{\rm o} = \sum_{{\rm odd}\,j} (2j+1)\exp[-xj(j+1)]\qquad{\rm
and}\qquad Z_{\rm p} = \sum_{{\rm even}\,j}
(2j+1)\exp[-xj(j+1)]\,,
\end{equation}
and $x\equiv \Theta_{\rm rot}/T = 85.4\,{\rm K}/T$
\citep{kittel58}. The dependence of $\gamma$ on temperature $T$ is
shown in Figure \ref{fig:gamma}.

Altogether, then, we have a system of 17 equations
[(\ref{eqn:nzcont}) - (\ref{eqn:fluxcont}), (\ref{eqn:ptensor}),
and (\ref{eqn:eos})], which contain 21 unknowns ($\dr{n}$, $P_{\rm
n}$, $\inteng$, $\bb{E}$, $\bb{B}$, $\bb{j}$, $\psi$, $\vv{n}$,
$\vv{e}$, $\vv{i}$, $\vv{g_-}$, $\vv{g_+}$, $\vv{g_0}$, $\dr{e}$,
$\dr{i}$, $\dr{g_-}$, $\dr{g_+}$, $\dr{g_0}$, $\mc{E}$, $\mcb{F}$,
$\msb{P}$). To close the system, the densities of electrons, ions,
and charged grains ($\nr{e}$, $\nr{i}$, $\nr{g_-}$, and
$\nr{g_+}$) are calculated from the equilibrium chemical model
described below.

\section{The Chemical Model}\label{section:chem}

\subsection{Ionization Rate}\label{subsection:ionization}

The rate of ionization per unit volume is given by $\zeta\nr{H_2}$
and is (in principle) due to the following ionization sources:
ultraviolet radiation, cosmic rays, radioactivities, and thermal
ionization. The presence of molecules in molecular clouds implies
low levels of UV radiation, so it is usually neglected. UV
radiation was included by \citet{cm95} in their numerical
simulations of core formation and evolution. They found that UV
ionization dominates cosmic-ray ionization for visual extinctions
$A_{\rm V} \lesssim 10$ and can increase the degree of ionization
in the envelope by at least two orders of magnitude (see also
McKee 1989). The increase in ionization was found to {\em speed
up} core collapse by approximately 30\% because the central
gravitational field of a flattened cloud is stronger (i.e., less
diluted by the mass of the envelope) when matter in the envelope
is held farther away from the forming core. Once dynamical
contraction ensues, it was found that UV radiation has little
effect on the evolution of central quantities and therefore it is
usually neglected. For numerical reasons, however, we add to the
electron and ion number densities the second term in equation (4h)
of \citet{fm92} ($=467.64 \, \nr{H_2}^{-2}$ cm$^{-3}$) so as to maintain
a relatively large degree of ionization ($\sim 10^{-5} - 10^{-6}$)
(and therefore negligible ambipolar diffusion) in the low-density
($\nr{H_2}\lesssim 10^3$ cm$^{-3}$) cloud envelope. This term
qualitatively mimics the effect of cloud envelope penetration by
UV photons, and has negligible quantitative effect on the
formation and evolution of the core.

Cosmic rays, on the other hand, with typical energies of 100 MeV
are able to penetrate deeper into molecular clouds. \citet{un80}
have investigated the ionization due to a spectrum of cosmic rays
at various energies. They found that the cosmic-ray ionization
rate was well described by the following relation:
\begin{equation}
\zeta_{\rm CR} = \zeta_0 \exp(-\Sigma_{\rm H_2}/96\,{\rm
g\,cm}^{-2})\,,
\end{equation}
where $\Sigma_{\rm H_2}$ is the column density of H$_2$ separating
the point in question from the exterior of the cloud and
$\zeta_0=5\times 10^{-17}$ s$^{-1}$ is the canonical unshielded
cosmic-ray ionization rate \citep{spitzer78}. \citet{tm07b} found
that, when a typical core's central density exceeds $\simeq
10^{12}$ cm$^{-3}$, cosmic rays are shielded and an abrupt
decrease in ionization occurs.

Once the core is shielded from high-energy cosmic rays, the
dominant source of ionization is radioactive decay of $^{40}$K or
$^{26}$Al. The isotope $^{40}$K is the most common radionuclide
invoked, due to its long half-life of 1.25 Gyr and its ubiquity in
nature (0.012\% of terrestrial potassium is $^{40}$K). The density
of potassium in the interstellar medium ($2.70\times
10^{-7}\,\nr{H_2}$) and the energy of the beta particle emitted as
$^{40}$K decays, 1.31 MeV, are used as inputs to calculate the
ionization rate (e.g., see Glassgold 1995): $\zeta_{40} =
2.43\times 10^{-23}$ s$^{-1}$. \citet{cj78} have suggested that
$^{26}$Al may have been a much more potent ionizer than $^{40}$K.
Performing a similar calculation for $^{26}$Al, one finds
$\zeta_{26} = 1.94\times 10^{-19}$ s$^{-1}$, with the fraction of
aluminum in the isotope $^{26}$Al inferred to have existed in the
solar nebula being $5\times 10^{-5}$ \citep{cl87}. Although
$^{26}$Al is four orders of magnitude more potent an ionizer than
$^{40}$K, its short half-life (0.716 Myr) makes it relevant only
if the initial mass-to-flux ratio of the parent cloud is close to
critical, so that the evolution is rapid enough to retain an
adequate amount of this radionuclide. $^{26}$Al can also become
important if the core happens to get enriched because of a nearby
Supernova explosion.

Finally, at temperatures on the order of 1000 K or higher,
collisions between molecules are energetic enough to ionize those
atoms with low ionization potentials, of which potassium and
sodium are the most abundant. The abundance of sodium in the
interstellar medium is greater than that of potassium (by a factor
$\simeq 14$; Lequeux 1975), but the lower threshold of potassium
(4.34 eV vs. 5.13 eV for sodium) makes it the dominant ion. The
ionization occurs at a rate \citep{pm65} given by
\begin{equation}
\frac{d}{dt}(\nr{K^+}) = 4.1\times 10^{-15}\nr{H_2}
\nr{K^0}\left(\frac{T}{1000\,{\rm K}
}\right)^{1/2}\exp\left(-\frac{5.04\times 10^4\,{\rm
K}}{T}\right)\,{\rm cm}^3\,{\rm s}^{-1}\,.
\end{equation}
Because this process relies on collisions between two species, it
is not expressed in terms of a quantity $\zeta$.

\subsection{Grain Size Distribution}\label{subsection:grainsize}

Since the dust opacity, the conductivity of the gas, and the
collision rates (see below) all depend on the (local) grain
surface area, it is necessary to investigate the effect of a grain
size distribution. The initial size distributions adopted here are
a uniform distribution and the standard ``MRN" distribution of
interstellar dust \citep{mrn77}. In both cases, the density of the
solid material of each grain is taken to be $\dr{S} = 2.3$ g
cm$^{-3}$, the average density of silicates. For the uniform
distribution, a fiducial grain size $a_0 = 0.0375$ $\mu$m is
used and the total mass density of dust $\rho_{\rm g,tot} =
0.01\rho_{\rm n,tot}$. For the MRN distribution, the number
density of spherical dust grains with radii between $a$ and $a +
da$ is
\begin{equation}\label{eqn:mrn}
dn_{\rm g,tot}(a) = N_{\rm MRN} a^{-3.5} da\,.
\end{equation}
The distribution is truncated at a lower grain radius $a_{\rm
min}$ and an upper grain radius $a_{\rm max}$.
The coefficient $N_{\rm MRN}$ is proportional to the dust-to-gas
mass ratio in the cloud. Note that most of the grain surface area
is contributed by small grains, because of their overwhelming
abundance.

The grains are binned according to size and charge and treated as
separate grain species. Each size bin represents a subset of the
original distribution of grains, those with radii between $a_{\rm
lower}$ and $a_{\rm upper}$. The subset of grains in the
$\alpha$th ($\alpha=1,2,\dots,N$) size bin is replaced by a number
density $n_{\rm g}^\alpha$ of grains with identical radii,
$a_\alpha$. The total number of grains and the total surface area
of grains in the size bin are constrained to match the total
number and surface area of original grains incorporated into the
size bin. Hence
\begin{equation}
\renewcommand{\theequation}{\arabic{equation}a,b}
n_{\rm g}^\alpha = \int_{a_{\rm lower}}^{a_{\rm upper}}
\frac{dn_{\rm g,tot}}{da} da\,,\qquad n_{\rm g}^\alpha a^2_\alpha
= \int_{a_{\rm lower}}^{a_{\rm upper}} a^2 \frac{dn_{\rm
g,tot}}{da} da\,.
\end{equation}
Applying these relations to the MRN grain size distribution,
equation (\ref{eqn:mrn}), if there are $N$ size bins, then the
$\alpha$th bin is characterized by grains of number density and
radii as follows:
\begin{equation}
\renewcommand{\theequation}{\arabic{equation}a,b}
n_{\rm g}^\alpha = n_{\rm g,tot}\, \xi^{2.5(\alpha-1)/N}
\left(\frac{1-\xi^{2.5/N}}{1-\xi^{2.5}}\right)\,,\qquad a_\alpha =
a_{\rm min}\,
\xi^{-(\alpha-1)/N}\left[5\left(\frac{1-\xi^{0.5/N}}{1-\xi^{2.5/N}}\right)\right]^{1/2}\,.
\end{equation}
The ratio of the lower and upper radii of the distribution is
denoted by $\xi\equiv a_{\rm min}/a_{\rm max}$. The total number
density of dust, $n_{\rm g,tot}$, is determined by constraining
the total grain mass density in the size distribution to be
$\rho_{\rm g,tot}$:
\begin{equation}\label{eqn:ngtotmrn}
n_{\rm g,tot} = \left(\frac{\rho_{\rm
g,tot}}{\frac{4}{3}\pi\dr{S}a^3_{\rm
min}}\right)\left[\frac{1}{5}\left(\frac{1-\xi^{2.5}}{1-\xi^{0.5}}\right)\right]
\xi^{0.5}\,.
\end{equation}
The lower and upper cutoffs to the size distribution are chosen to
be $a_{\rm min} = 0.0181$ $\mu$m and $a_{\rm max} = 0.9049$
$\mu$m, respectively. In equation (\ref{eqn:ngtotmrn}), the total
mass density of dust in the system, $\rho_{\rm g,tot}$, is chosen
in such a way that the total grain surface area in the size
distribution is equal to that in the fiducial case of a single
grain size $a_0$. This constraint demands that $\rho_{\rm g,tot}$
be increased by a factor $(a_{\rm min}/a_0)\,\xi^{-0.5}$ over the
fiducial value of $\rho_{\rm g,tot}$. Only in this way can the
effect of a size distribution, as distinct from just the surface
area of grains, be determined. With the fiducial values
$a_0=0.0375$ $\mu$m and $\rho_{\rm g,tot} = 0.01 \rho_{\rm n,tot}$
for the single grain case, $\rho_{\rm g,tot}=0.0341 \rho_{\rm
n,tot}$ for the case of a size distribution. Empirically, it was
found that a minimum of five size bins of grain radii were
required for convergence of 1\%. Since each size grain can be
found in one of three possible charge states ($-e$, 0, and $+e$),
a total of fifteen grain species are considered.

While we do not consider grain growth (and therefore fix the
number of grains within each size bin), we do expect the grain
size distribution to evolve spatially within the star-forming
cloud. Ambipolar diffusion can alter a grain size distribution by
acting more effectively on the larger grains, causing a spatial
segregation of grain sizes that leaves the smaller grains behind
in the cloud envelope. The result is a deficit of small grains
($a\lesssim 10^{-5}$ cm) in the cloud core. In fact, \citet{cm96}
show how observations of grain abundances in the core and envelope
of a molecular cloud can, at least in principle, be used to
determine the initial mass-to-flux ratio of the cloud.

\subsection{Chemical Network}\label{subsection:chem}

We use a chemical equilibrium network accounting for electrons
(subscript e); molecular ions such as HCO$^+$ (subscript m$^+$);
neutral metal atoms (subscript A$^0$) and atomic ions (subscript
A$^+$) of Mg, Na, and K; singly positively-charged grains
(subscript g$_+$); singly negatively-charged grains (subscript
g$_-$); and, neutral grains (subscript g$_0$). Multiply
negatively- (positively-)charged grains may be neglected, because
a singly negatively- (positively-)charged grain repels electrons
(ions) thereby decreasing the rate of capture by the factor
$\exp(-e^2/ak_{\rm B}T)$ \citep{spitzer41}. The equilibrium
assumption is accurate provided that the dynamical timescales of
interest are sufficiently longer than the chemical-reaction
timescales. This is always the case for the density regime
considered here. The relevant reactions are given below and
explained briefly.

The production of molecular ions (such as HCO$^+$) is balanced by
their destruction through charge-exchange reactions with atomic
ions, by dissociative recombinations (collisions with electrons),
or by collisions with and neutralization on the surfaces of
grains:
\begin{equation}\label{eqn:mpchem}
\zeta\nr{H_2} = \nr{m^+}\nr{A^0}\beta + \nr{m^+}\nr{e}\alpha_{\rm
dr} + \sum_\alpha\nr{m^+}\ngs{-}{\alpha}\alpha_{{\rm
m^+g}^\alpha_-} + \sum_\alpha\nr{m^+}\ngs{0}{\alpha}\alpha_{{\rm
m^+g}^\alpha_0}\,.
\end{equation}
The index $\alpha$ denotes a grain size bin, and the sum is over
all the size bins, which are treated as independent grain species.
The production of atomic ions by charge-exchange reactions is
balanced by radiative recombinations and by collisions with
grains:
\begin{equation}\label{eqn:apchem}
\nr{m^+}\nr{A^0}\beta = \nr{A^+}\nr{e}\alpha_{\rm rr} +
\sum_\alpha\nr{A^+}\ngs{-}{\alpha}\alpha_{{\rm A^+g}^\alpha_-} +
\sum_\alpha\nr{A^+}\ngs{0}{\alpha}\alpha_{{\rm A^+g}^\alpha_0}\,.
\end{equation}
If the atomic ion in question is K$^+$, there is the additional
source term due to thermal ionization of potassium atoms,
$\nr{K^0}\nr{H_2}\alpha_{\rm K^0H_2}$. Positively-charged grains
are formed by the collisions of ions and neutral grains and by
charge exchange between grains; they are destroyed by collisions
with electrons, collisions with negative grains, and by charge
exchange with neutral grains:
\begin{equation}\label{eqn:gpchem}
\nr{m^+}\ngs{0}{\alpha}\alpha_{{\rm m^+g}^\alpha_0} +
\nr{A^+}\ngs{0}{\alpha}\alpha_{{\rm A^+g}^\alpha_0} +
\sum_{\alpha'}\ngs{+}{\alpha'}\ngs{0}{\alpha}\alpha_{{\rm
g}^{\alpha'}_+{\rm g}^\alpha_0} =
\nr{e}\ngs{+}{\alpha}\alpha_{{\rm eg}^\alpha_+} +
\sum_{\alpha'}\ngs{+}{\alpha}\ngs{-}{\alpha'}\alpha_{{\rm
g}^\alpha_+{\rm g}^{\alpha'}_-} +
\sum_{\alpha'}\ngs{+}{\alpha}\ngs{0}{\alpha'}\alpha_{{\rm
g}^\alpha_+{\rm g}^{\alpha'}_0}\,.
\end{equation}
Here the index $\alpha'$ runs over all the grain size bins,
independently of the index $\alpha$. Negatively-charged grains are
formed by the collisions of electrons and neutral grains and by
charge exchange between grains, and are destroyed by collisions
with ions, collisions with positive grains, and by charge exchange
with neutral grains:
\begin{equation}\label{eqn:gmchem}
\nr{e}\ngs{0}{\alpha}\alpha_{{\rm eg}^\alpha_0} +
\sum_{\alpha'}\ngs{-}{\alpha'}\ngs{0}{\alpha}\alpha_{{\rm
g}^{\alpha'}_-{\rm g}^\alpha_0} =
\nr{m^+}\ngs{-}{\alpha}\alpha_{{\rm m^+g}^\alpha_-} +
\nr{A^+}\ngs{-}{\alpha}\alpha_{{\rm A^+g}^\alpha_-} +
\sum_{\alpha'}\ngs{+}{\alpha'}\ngs{-}{\alpha}\alpha_{{\rm
g}^{\alpha'}_+{\rm g}^\alpha_-} +
\sum_{\alpha'}\ngs{-}{\alpha}\ngs{0}{\alpha'}\alpha_{{\rm
g}^\alpha_-{\rm g}^{\alpha'}_0} \,.
\end{equation}
We close this set of equations with constraints on the total
number of grains in a given size bin,
\begin{equation}\label{eqn:gzchem}
\ngs{+}{\alpha} + \ngs{0}{\alpha} + \ngs{-}{\alpha} = n_{{\rm
g}^\alpha}\,,
\end{equation}
and the total number of an atomic species (neutral +
positively-charged),
\begin{equation}\label{eqn:azchem}
\nr{A^0} + \nr{A^+} = n_{\rm A}\,,
\end{equation}
and with charge neutrality:
\begin{equation}\label{eqn:emchem}
\nr{m^+} + \nr{A^+} - \nr{e} +
\sum_\alpha\bigl(\ngs{+}{\alpha}-\ngs{-}{\alpha}\bigr) = 0\,.
\end{equation}
This system of equations [(\ref{eqn:mpchem}) - (\ref{eqn:emchem})]
is solved numerically via Gauss-Jordan elimination on the matrix
equation derived by applying the Newton-Raphson iteration method.
The rate coefficients in equations (\ref{eqn:mpchem}) -
(\ref{eqn:emchem}) are given in Appendix \ref{app:ratecoeffs}.

The mass of molecular ions is taken to be that of HCO$^+$, $m_{\rm
m^+} = 29\,m_{\rm p}$, while for the atomic ions an average value
$m_{\rm A^+} = 23.5\,m_{\rm p}$, between the mass of Na ($m_{\rm
Na^+} = 23\,m_{\rm p}$) and the mass of Mg ($m_{\rm
Mg^+}=24\,m_{\rm p}$), is used. Since the ion masses are all
comparable, the fact that different ionic species dominate in
different density regimes does not affect the evolution of the
cloud cores. The total number of atomic ions is fixed at
$2.05\times 10^{-6}\,\nr{n}$ \citep{morton74,snow76}.

\section{Magnetic Flux Loss and Electrical Resistivity}\label{section:fluxloss}

The force equations [(\ref{eqn:emforce}) - (\ref{eqn:gzforce})]
and the induction equation (\ref{eqn:faraday}) are not written in
the most convenient form for our purposes (see
\S~\ref{section:zeusmp} below). A useful simplification can be
made, which amounts to a generalized version of Ohm's law; namely,
we replace equations (\ref{eqn:emforce}) - (\ref{eqn:gzforce})
with a modified form of equation (\ref{eqn:faraday}). This
auspiciously eliminates 5 variables ($\vv{e}$, $\vv{i}$,
$\vv{g_0}$, $\vv{g_-}$, and $\vv{g_+}$), but not without a cost.
The ensuing algebra is messy, and much of it is deferred to
Appendix \ref{app:ohmslaw}. Here, we outline the simplification
and highlight some results suitable for the present discussion.

\subsection{Resistivity of a Magnetic Gas}\label{subsection:resistivity}

The rate of change of magnetic flux across a given surface $S$,
comoving with a fluid with velocity $\bb{v}$, is given by
\begin{equation}
\frac{d\Phi_{\rm B}}{dt} = \int_S\left[\D{t}{\bb{B}} -
\del\btimes(\bb{v}\btimes\bb{B})\right]\bcdot d\bb{S}\,.
\end{equation}
Using Faraday's law (\ref{eqn:faraday}), the integrand can be
rewritten as
\begin{equation}\label{eqn:fluxfaraday}
\frac{d\Phi_{\rm B}}{dt} = -c \int_S \del\btimes\left(\bb{E} +
\frac{\bb{v}}{c}\btimes\bb{B}\right)\bcdot d\bb{S}\,,
\end{equation}
and the current density can be calculated from
\begin{equation}\label{eqn:johm}
\bb{j} =
\bb{\sigma}\left(\bb{E}+\frac{\bb{v}}{c}\btimes\bb{B}\right)\,.
\end{equation}
The quantity $\bb{v}$ is the velocity of the fluid, which for a
weakly-ionized gas is essentially that of the neutrals $\vv{n}$,
and $\bb{\sigma}$ is the conductivity tensor. The presence of a
magnetic field introduces an anisotropy in the equations, which is
the reason for which the conductivity must be described by a
tensor. If we take the 3-direction to lie along the magnetic
field, the conductivity tensor has the following representation
\citep{parks91}:
\begin{equation}
\bb{\sigma} =
\begin{pmatrix}
\sigma_\perp & -\sigma_{\rm H} & 0 \\
\sigma_{\rm H} & \sigma_\perp & 0 \\
0 & 0 & \sigma_{||}
\end{pmatrix}\,.
\end{equation}
As $B\rightarrow 0$, the tensor must reduce to an isotropic form;
i.e., $\sigma_{\rm H}\rightarrow 0$ and
$\sigma_\perp\rightarrow\sigma_{\rm ||}$. Because magnetic forces
vanish along the magnetic field, $\sigma_{||}$ must be independent
of the magnetic field strength.

Equation (\ref{eqn:johm}) may be inverted to obtain the
electric field, in which case a resistivity tensor $\bb{\eta}$ is
defined by
\begin{equation}\label{eqn:Eohm}
\bb{E} + \frac{\vv{n}}{c}\btimes\bb{B} = \bb{\eta}\bb{j}\,,
\end{equation}
where, in the same representation as $\bb{\sigma}$ written above,
\begin{equation}
\bb{\eta} =
\begin{pmatrix}
\eta_\perp & \eta_{\rm H} & 0 \\
-\eta_{\rm H} & \eta_\perp & 0 \\
0 & 0 & \eta_{||}
\end{pmatrix}\,,
\end{equation}
and
\begin{equation}\label{eqn:etadef}
\renewcommand{\theequation}{\arabic{equation}a,b,c}
\eta_{||} = \frac{1}{\sigma_{||}}\,,\qquad \eta_\perp =
\frac{\sigma_\perp}{\sigma_\perp^2 + \sigma_{\rm H}^2}\,,\qquad
\eta_{\rm H} = \frac{\sigma_{\rm H}}{\sigma_\perp^2+\sigma_{\rm
H}^2}\,.
\end{equation}
The flux-freezing approximation corresponds to the limit
$\bb{\eta}\rightarrow 0$.

If we write the current density $\bb{j}$ in component form, it
follows that we may write equation (\ref{eqn:Eohm}) as
\begin{equation}
\bb{E} + \frac{\vv{n}}{c}\btimes\bb{B} = \eta_{||}\bb{j}_{||} +
\eta_\perp\bb{j}_\perp + \eta_{\rm H}\bb{j}\btimes\eb\,,
\end{equation}
where $\bb{j}_{||}$ and $\bb{j}_\perp$ are the components of the
current density parallel and perpendicular to the magnetic field,
respectively, and $\eb$ is a unit vector along the magnetic field.
This relation between the electric field and the current density
can be substituted in equation (\ref{eqn:fluxfaraday}) to find
that
\begin{equation}
\frac{d\Phi_{\rm B}}{dt} = -c\int_S
\del\btimes\bigl(\eta_{||}\bb{j}_{||} + \eta_\perp\bb{j}_\perp +
\eta_{\rm H}\bb{j}\btimes\eb\bigr)\bcdot d\bb{S}\,.
\end{equation}
This is the general form of the equation describing the loss of
magnetic flux from a parcel of neutral gas, written entirely in
terms of the components of the resistivity and current density
tensors.

For our model cloud, we have assumed axisymmetry and neglected
rotation. In this case, the magnetic field is purely poloidal and
the current density is purely toroidal by Ampere's law
(\ref{eqn:ampere}). This geometry implies that the only nonvanishing
component of the current density is the component perpendicular to
the magnetic field, $\bb{j}=\bb{j}_\perp$. The evolution of the
poloidal magnetic flux in the neutrals' reference frame is then
given by
\begin{equation}\label{eqn:fluxfaraday2}
\frac{d\Phi_{\rm B}}{dt} = -c\int_S
\del\btimes\bigl(\eta_\perp\bb{j}_\perp\bigr)\bcdot d\bb{S}\,.
\end{equation}
The equivalent equation governing the evolution of the poloidal
magnetic field is
\begin{equation}\label{eqn:faraday2}
\D{t}{\bb{B}} = \del\btimes\bigl(\vv{n}\btimes\bb{B}\bigr) -
c\del\btimes\bigl(\eta_\perp\bb{j}_\perp\bigr)\,.
\end{equation}
Equation (\ref{eqn:fluxfaraday2}) describes the evolution of
magnetic flux in the neutrals' reference frame due to the motion
of charges at right angles to the magnetic field and includes the
effects of both ambipolar diffusion and Ohmic dissipation. The
rate at which magnetic flux is lost equals the sum of the rates
due to each process. Therefore, $\eta_\perp$ can itself be written
as the sum of two components, one related to ambipolar diffusion
(subscript AD) and the other to Ohmic dissipation (subscript OD):
\begin{equation}
\eta_\perp = \eta_{\perp{\rm AD}} + \eta_{\perp{\rm OD}}\,.
\end{equation}
The issue of how to separate the resistivity $\eta_\perp$ into its
two components is discussed, for example, in \citet{nu86a,nu86b},
\citet{gr92}, and \citet{dm01}. We quote the result here:
\begin{equation}\label{eqn:etasep}
\renewcommand{\theequation}{\arabic{equation}a,b,c,d}
\eta_{||{\rm OD}} = \eta_{||}\,,\qquad \eta_{\perp{\rm OD}} =
\eta_{||}\,,\qquad \eta_{||{\rm AD}} = 0\,,\qquad \eta_{\perp{\rm
AD}} = \eta_\perp - \eta_{||}\,.
\end{equation}
We now derive expressions for the resistivities from first
principles.

\subsection{Generalized Ohm's Law}\label{subsection:ohmslaw}

We outline the derivation of a generalized Ohm's law, taking into
account both elastic and inelastic collisions between neutrals,
ions, electrons, and charged and neutral grains. We begin by
writing down the force equation for the charged species $s$:
\begin{equation}\label{eqn:sforce}
0 = n_s q_s \left(\bb{E}+\frac{\vs{s}}{c}\btimes\bb{B}\right) +
\frac{\rho_s}{\tausn{s}}(\vv{n}-\vs{s}) +
\frac{\dr{g_0}}{\tsinel{s}}(\vv{g_0}-\vs{s})\,.
\end{equation}
The subscript $s$ runs over all the charged species, taking on the
values $s$ = i, e, g$_+$, and g$_-$. Although we employ a grain
size distribution, we consider only a single grain size in what
follows for ease of presentation; a discussion of the consequences
of a grain size distribution is deferred to Appendix
\ref{app:grainsize}. The charge $q_s$ of species $s$ carries an
algebraic sign (e.g., it is negative for electrons). We write
$\tsinel{s}$ to represent the timescale for species $s$ to be
created by or take part in any inelastic collision. For example,
\begin{equation}\label{eqn:tgpinel}
\trinel{g_+} = \left[\frac{1}{\trinel{g_0i}} +
\frac{1}{\trinel{g_0g_+}} +
\frac{\dr{g_+}}{\dr{g_0}}\left(\frac{1}{\trinel{g_+e}} +
\frac{1}{\trinel{g_+g_-}} +
\frac{1}{\trinel{g_+g_0}}\right)\right]^{-1}
\end{equation}
is the timescale for a neutral grain to participate in {\em any}
inelastic reaction involving conversion between positive and
neutral grains. The first two terms represent the production of
positive grains due to charge exchange between neutral grains and
ions and between neutral grains and positive grains, respectively;
the next two terms represent the conversion of positive grains to
neutral grains via neutralization with electrons and negative
grains, respectively; and, the final term represents the
conversion of positive grains to neutral grains via charge
exchange. Since these are processes occurring in parallel, the
reciprocals of their respective collision times are added to
obtain the net collision time. Similarly,
\begin{equation}\label{eqn:tgminel}
\trinel{g_-} = \left[\frac{1}{\trinel{g_0e}} +
\frac{1}{\trinel{g_0g_-}} +
\frac{\dr{g_-}}{\dr{g_0}}\left(\frac{1}{\trinel{g_-i}} +
\frac{1}{\trinel{g_-g_+}} +
\frac{1}{\trinel{g_-g_0}}\right)\right]^{-1}
\end{equation}
is the timescale for a neutral grain to be involved in {\em any}
inelastic reaction involving conversion between negative and
neutral grains. The force equation for the neutral grains is
\begin{equation}\label{eqn:gzforce2}
0 = \frac{\dr{g_0}}{\taurn{g_0}}(\vv{n}-\vv{g_0}) + \sum_k
\frac{\dr{g_0}}{\tsinel{k}}(\vs{k}-\vv{g_0})\,,
\end{equation}
where the index $k$ runs over all the charged species.

We eliminate the velocity $\vs{s}$ of species $s$ in favor of a
new velocity, $\ws{s}$, which is the velocity of species $s$ with
respect to the neutral gas ($\ws{s}\equiv\vs{s}-\vv{n}$). In
addition, we define $\bb{E}_{\rm n}$ as the electric field in the
frame of reference of the neutral gas ($\bb{E}_{\rm n} \equiv
\bb{E} + \vv{n}\btimes\bb{B}/c$). Equations (\ref{eqn:sforce}) and
(\ref{eqn:gzforce2}) then become, respectively,
\begin{equation}\label{eqn:sforce2}
0 =
\frac{\omega_s\tausn{s}}{1+\varrho_s}\left(\frac{c}{B}\bb{E}_{\rm
n}+\ws{s}\btimes\eb\right) - \ws{s} +
\frac{\varrho_s}{1+\varrho_s}\ww{g_0}\,,
\end{equation}
\begin{equation}\label{eqn:gzforce3}
0 = \ww{g_0} - \sum_k \frac{\tau_0}{\tsinel{k}}\ws{k}\,,
\end{equation}
where we have introduced the cyclotron frequency of species $s$,
$\omega_s = q_s B / m_s c$, and have defined $\varrho_s$ and
$\tau_0$ by
\begin{equation}\label{eqn:tauz}
\renewcommand{\theequation}{\arabic{equation}a,b}
\varrho_s =
\frac{\dr{g_0}}{\rho_s}\frac{\tausn{s}}{\tsinel{s}}\qquad{\rm
and}\qquad \frac{1}{\tau_0} = \frac{1}{\taurn{g_0}} + \sum_k
\frac{1}{\tsinel{k}}\,.
\end{equation}
Equations (\ref{eqn:sforce2}) and (\ref{eqn:gzforce3}) form the
set of equations to be solved.\footnote{The quantity $\varrho_s$
was written as $r_s$ in \citet{tm07a}. We have renamed it here to
avoid confusion with the cylindrical radial coordinate $r$.} The
species velocities (relative to the neutrals) $\ws{s}$ can be
expressed in terms of $\bb{E}_{\rm n}$ and then substituted in the
definition of the current density
\begin{equation}
\bb{j} = \sum_s n_s q_s \ws{s}\,,
\end{equation}
where we have used charge neutrality ($\sum_s n_s q_s = 0$). This
expression can then be inverted to find $\bb{E}_{\rm n}$ in terms
of $\bb{j}$, which defines the resistivity tensor. The magnetic
induction equation is then found by substitution into Faraday's
law of induction:
\begin{equation}
\D{t}{\bb{B}} - \del\btimes(\vv{n}\btimes\bb{B}) =
-c\del\btimes\bb{E}_{\rm n}\,.
\end{equation}
Using this approach, we derive an induction equation generalized
to include Ohmic dissipation, ambipolar diffusion, and the Hall
effect for a six-fluid system including both elastic and inelastic
collisions.

The ensuing calculation is tedious, and we defer the details to
Appendix \ref{app:ohmslaw}. Here we give only the final result:
\begin{equation}\label{eqn:ohmslawj}
\bb{j} = \sigma_{||}\bb{E}_{\rm n,||} + \sigma_\perp\bb{E}_{\rm
n,\perp} - \sigma_{\rm H}\bb{E}_{\rm n}\btimes\eb\,,
\end{equation}
where
\begin{equation}\label{eqn:sigmas}
\renewcommand{\theequation}{\arabic{equation}a,b,c}
\sigma_{||} = \sum_s \sigma_s (1-\varsigma_s)\,,\qquad
\sigma_\perp = \sum_s
\frac{\sigma_s(1-\varsigma_s)}{1+\omega^2_s\tausn{s}^2(1-\varphi_s)}\Upsilon_s(\varsigma)\,,\qquad
\sigma_{\rm H} =
-\sum_s\frac{\sigma_s\omega_s\tausn{s}(1-\varpi_s)}{1+\omega^2_s\tausn{s}^2(1-\varphi_s)}\Upsilon_s(\varpi)\,.
\end{equation}
The conductivity of species $s$ is given by $\sigma_s = n_s q^2_s
\tausn{s}/m_s$. The quantities  $\varsigma_s$, $\varpi_s$,
$\varphi_s$, and $\Upsilon_s$ are defined in Appendix
\ref{app:defs}; they represent the effects of inelastic collisions
on the conductivity of the gas. In the absence of inelastic
collisions, these formulae reduce to their standard form (e.g.,
see Parks 1991):
\begin{equation*}
\sigma_{||} \rightarrow \sum_s \sigma_s\,,\qquad\sigma_\perp
\rightarrow \sum_s
\frac{\sigma_s}{1+\omega^2_s\tausn{s}^2}\,,\qquad\sigma_{\rm H}
\rightarrow
-\sum_s\frac{\sigma_s\omega_s\tausn{s}}{1+\omega^2_s\tausn{s}^2}\,.
\end{equation*}
Equation (\ref{eqn:ohmslawj}) may be inverted to give
\begin{equation}\label{eqn:ohmslawE}
\bb{E}_{\rm n} = \eta_{||}\bb{j}_{||} + \eta_\perp\bb{j}_\perp +
\eta_{\rm H}\bb{j}\btimes\eb\,,
\end{equation}
with the resistivities $\eta_{||}$, $\eta_\perp$, and $\eta_{\rm
H}$ given by equations (\ref{eqn:etadef}).

\subsection{Attachment of Species to Magnetic Field Lines}\label{subsection:attachment}

It is possible to write the velocity of each species, $\vs{s}$, in
terms of the velocity of the neutrals, $\vv{n}$, and the velocity
of the field lines, $\vv{f}$, which is defined implicitly by
\begin{equation}
\bb{E} + \frac{\vv{f}}{c}\btimes\bb{B} = 0\,.
\end{equation}
The algebra and some intermediate results of interest are given in
Appendix \ref{app:velocities}; here, we quote the main result and
explain it physically:
\begin{subequations}
\begin{align}\label{eqn:velocitiesvf}
\vs{s,\perp} &= \vv{n,\perp} \frac{1}{\Theta_s+1} + \vv{f,\perp} \frac{\Theta_s}{\Theta_s+1} +(\vv{f}-\vv{n})\btimes\eb\,\Lambda_s\,, \\
\vs{s}\btimes\eb &= \vv{n}\btimes\eb \frac{1}{\Theta_s+1} +
\vv{f}\btimes\eb \frac{\Theta_s}{\Theta_s+1} -
(\vv{f,\perp}-\vv{n,\perp})\,\Lambda_s\,,
\end{align}
\end{subequations}
where the expressions for $\Theta_s$ and $\Lambda_s$ are given in
Appendix \ref{app:velocities}. The quantity $\Theta_s$ is the {\em
attachment parameter} (i.e., for $\Theta_s\gg 1$,
$\vs{s}\approx\vv{f}$ and species $s$ is attached to the field
lines, whereas for $\Theta_s\ll 1$, $\vs{s}\approx\vv{n}$ and
species $s$ is detached and comoves with the neutrals) --- see,
also, \citet{cm93}, \S~3.1.2. The function
$\Lambda_s$ quantifies the relation of one component of the
species velocity to its mutually perpendicular component of the
field line drift velocity, and essentially embodies Ampere's law.
Under the assumptions of this paper, the midplane velocities of
the charged species $s$, written in cylindrical coordinates
$(r,\phi,z)$, are
\begin{equation}
\renewcommand{\theequation}{\arabic{equation}a,b}
\vs{s,\phi}(r,z=0) = (\vv{n,r}-\vv{f,r})\,\Lambda_s\,,\qquad
\vs{s,{\rm r}}(r,z=0) = \vv{n,r}\frac{1}{\Theta_s+1} +
\vv{f,r}\frac{\Theta_s}{\Theta_s+1}\,.
\end{equation}
The first equation says that the charged species move in such a
way as to cause differential motion between the field lines and
the neutrals (Ampere's law). The second equation gives the radial
velocity of any charged species in terms of the the velocities of
the neutrals and of the field lines. These may be combined to
yield
\begin{equation}
\ws{s,{\rm r}} = \vs{s,{\rm r}} - \vv{n,r} =
-\frac{\Theta_s}{\Theta_s+1}\frac{\vs{s,\phi}}{\Lambda_s}\,.
\end{equation}
In other words, the radial drift between species $s$ and the
neutrals is directly proportional to the contribution of species
$s$ to the azimuthal current.

\subsection{Grain Continuity Equation}\label{subsection:graincont}

In the notation of Section \ref{subsection:ohmslaw}, the grain
continuity equation (\ref{eqn:gtcont}) may be written as
\begin{equation}
\D{t}{\dr{g}} + \del\bcdot(\dr{g}\vv{n}) =
-\del\bcdot(\dr{g_-}\ww{g_-}+\dr{g_0}\ww{g_0}+\dr{g_+}\ww{g_+})\,,
\end{equation}
where $\dr{g}=\dr{g_-}+\dr{g_0}+\dr{g_+}$ is the total grain
density. Eliminating $\ww{g_0}$ using equation
(\ref{eqn:gzforce3}), we find that
\begin{equation}
\D{t}{\dr{g}} + \del\bcdot(\dr{g}\vv{n}) =
-\del\bcdot\left[\dr{g_+}\ww{g_+}\left(1+\frac{\tau_0}{\taurn{g_+}}\varrho_{\rm
g_+}\right) +
\dr{g_-}\ww{g_-}\left(1+\frac{\tau_0}{\taurn{g_-}}\varrho_{\rm
g_-}\right)\right]\,.
\end{equation}
We may use equation (\ref{eqn:velocitiesj}) to eliminate the
differential velocities of the charged grain species to find,
after some manipulation,
\begin{equation}\label{eqn:gtcont2}
\D{t}{\dr{g}} + \del\bcdot(\dr{g}\vv{n}) =
-\del\bcdot\bigl(\eta_{\rm cont,||}\bb{j}_{||} + \eta_{\rm
cont,\perp}\bb{j}_\perp + \eta_{\rm
cont,H}\bb{j}\btimes\eb\bigr)\,,
\end{equation}
where the components of the grain-continuity resistivity tensor,
$\bb{\eta}_{\rm cont}$, are defined as
\begin{subequations}
\begin{equation}
\eta_{\rm cont,||} = \sum_{s={\rm g_+,g_-}}
\frac{m_s}{q_s}\left[\eta_{||}\sigma_{||,s}\left(1+\frac{\tau_0}{\tausn{s}}\varrho_s\right)\right]\,,
\end{equation}
\begin{equation}
\eta_{\rm cont,\perp} = \sum_{s={\rm g_+,g_-}}
\frac{m_s}{q_s}\left[\eta_\perp\sigma_{\perp,s}\left(1+\frac{\tau_0}{\tausn{s}}\varrho_s\right)-\eta_{\rm
H}\sigma_{{\rm
H},s}\left(1+\frac{\tau_0}{\tausn{s}}\varrho_s\right)\right]\,,
\end{equation}
\begin{equation}
\eta_{\rm cont,H} = \sum_{s={\rm g_+,g_-}}
\frac{m_s}{q_s}\left[\eta_{\rm
H}\sigma_{\perp,s}\left(1+\frac{\tau_0}{\tausn{s}}\varrho_s\right)-\eta_\perp\sigma_{{\rm
H},s}\left(1+\frac{\tau_0}{\tausn{s}}\varrho_s\right)\right]\,.
\end{equation}
\end{subequations}
These equations apply to all grain sizes separately. Under the
assumptions in this work, $\bb{j}_{||} = 0$ and
$\del\bcdot\bb{j}_\perp=0$ by axisymmetry. Equation
(\ref{eqn:gtcont2}) then becomes
\begin{equation}\label{eqn:gtcont3}
\D{t}{\dr{g}} + \del\bcdot(\dr{g}\vv{n}) =
-\del\bcdot\bigl(\eta_{\rm cont,H}\bb{j}\btimes\eb\bigr)\,.
\end{equation}
Note that if $\eta_{\rm cont,H} = 0$, a quantitative
implementation of flux-freezing, the grain species are advected
with the neutrals, as expected.

\subsection{Joule Heating}\label{subsection:joule}

The rate $\Gamma_{\rm diff}$ at which collisions dissipate kinetic
energy as heat per unit volume (in the reference frame of the
neutrals) may be calculated by taking the dot product of equation
(\ref{eqn:sforce2}) with $\ws{s}$ and using equation
(\ref{eqn:gzforce2}):
\begin{equation}
\Gamma_{\rm diff} = \left[\sum_s
(1+\varrho_s)\frac{\rho_s}{\tausn{s}}|\ws{s}|^2\right] -
\left(\sum_s\frac{\dr{g_0}}{\tsinel{s}}\ws{s}\right)\bcdot\left(\sum_k\frac{\tau_0}{\tsinel{k}}\ws{k}\right)\,,
\end{equation}
where the summation indices $s$ and $k$ run, as usual, over all
charged species (including charged grains of different sizes if a
grain size distribution is considered). Using equation
(\ref{eqn:velocitiesj}) to eliminate the velocities in favor of
the current density, we find after much simplification
\begin{align}\label{eqn:joule}
\Gamma_{\rm diff} &= \eta_{||}|\bb{j}_{||}|^2\left[\sum_s
\left(\frac{\sigma_{||,s}}{\sqrt{\sigma_s\sigma_{||}}}
\sqrt{1+\varrho_s}\right)^2 - \left(\sum_s
\frac{\sigma_{||,s}}{\sqrt{\sigma_s\sigma_{||}}}
\sqrt{\frac{\varrho_s\tau_0}{\tsinel{s}}}\right)^2\right] \nonumber\\
&+ \eta_{\perp}|\bb{j}_{\perp}|^2\left[\sum_s
\left(\frac{\sigma_{\perp,s}}{\sqrt{\sigma_s\sigma_{\perp}}}
\sqrt{1+\varrho_s}\right)^2 - \left(\sum_s \frac{\sigma_{\perp,s}}
{\sqrt{\sigma_s\sigma_{\perp}}}\sqrt{\frac{\varrho_s\tau_0}{\tsinel{s}}}\right)^2\right] \nonumber\\
&+ \eta_{\rm
H}|\bb{j}_{\perp}|^2\left[\sum_s\left(\frac{\sigma_{{\rm
H},s}}{\sqrt{\sigma_s\sigma_{\rm H}}}\sqrt{1+\varrho_s}\right)^2 -
\left(\sum_s \frac{\sigma_{{\rm H},s}}{\sqrt{\sigma_s\sigma_{\rm
H}}}\sqrt{\frac{\varrho_s\tau_0}{\tsinel{s}}}\right)^2\right] \,.
 \end{align}
In the limit where inelastic collisions are negligible relative to
elastic collisions (i.e., $\varrho_s\rightarrow 0$), this equation
reduces to the usual expression
\begin{equation*}
\Gamma_{\rm diff}\rightarrow \eta_{||}|\bb{j}_{||}|^2 +
\eta_\perp|\bb{j}_\perp|^2 = \eta_{\rm OD}|\bb{j}|^2 + \eta_{\rm
AD}|\bb{j}_\perp|^2\,.
\end{equation*}
In the last step, we have used equations (\ref{eqn:etasep}) to
separate the contributions of Ohmic dissipation and ambipolar
diffusion to the heating rate. Ohmic dissipation affects the total
current density, whereas ambipolar diffusion affects only the
perpendicular component of the current density.

\section{Radiative Transfer}\label{section:radtrans}

\subsection{The Flux-Limited Diffusion Approximation}\label{subsection:FLD}

Computing a formal solution of the full angle-frequency dependent
non-LTE radiative transfer equation in a multidimensional
numerical algorithm is a prohibitive task. Even if a rigorous yet
tractable algorithm were developed to this end, the computational
expense involved would prevent a solution in any reasonable amount
of time. In fact, the sophisticated numerical code described in
\citet{smn92} designed to solve this problem with as few
approximations as possible never saw public release. The FLD
approximation is an attractive method for handling transport
phenomena that is relatively easy to implement, robust, and
inexpensive. It has the advantage over other diffusive
approximations in that it preserves causality in regions where
significant spatial variation can occur over distances smaller
than a mean free path. For example, the Eddington approximation
consists of assuming the radiation field is everywhere isotropic,
an assumption that is violated in the optically-thin limit where
the radiation becomes streaming \citep{mm84}.

The fundamental assumption of FLD is that the specific intensity
is a slowly varying function of space and time. This is certainly
valid in the diffusion and streaming limits (at least in one
dimension); one hopes that it is approximately true in
intermediate situations (and in multi-dimensions). Given this
assumption, \citet{lp81} showed that the radiation flux can be
expressed in the form of Fick's law of diffusion,
\begin{equation}
\mcb{F} = -\mc{D}_{\rm FLD}\del\mc{E}\,,
\end{equation}
where the diffusion coefficient $\mc{D}_{\rm FLD}$ can be written
as
\begin{equation}
\mc{D}_{\rm FLD} = \frac{c\lambda_{\rm FLD}}{\chi_{\mc{F}}}\,.
\end{equation}
The dimensionless function $\lambda_{\rm FLD} = \lambda_{\rm
FLD}(\mc{E})$ is called the flux limiter. Similarly, in FLD theory
the radiation pressure tensor can be expressed in terms of the
radiation energy density via
\begin{equation}
\msb{P} = \msb{f}\mc{E}\,,
\end{equation}
where the components of the Eddington tensor $\msb{f}$ are given
by
\begin{equation}
\msb{f} = \frac{1}{2}(1-f)\msb{I} +
\frac{1}{2}(3f-1)\hat{\bb{n}}\hat{\bb{n}}\,,
\end{equation}
where $\hat{\bb{n}}=\del\mc{E}/|\del\mc{E}|$ is the normalized
gradient of $\mc{E}$ and the dimensionless function $f=f(\mc{E})$
is called the Eddington factor \citep{ts01}. The flux limiter
$\lambda_{\rm FLD}$ and Eddington factor $f$ are related through
implicit constraints between the moments $\mcb{F}$ and $\msb{P}$,
so that
\begin{equation}
f = \lambda_{\rm FLD} + \lambda^2_{\rm FLD}\mc{R}^2\,,
\end{equation}
where $\mc{R}$ is the dimensionless quantity $\mc{R} =
|\del\mc{E}|/\chif\mc{E}$. We have chosen the flux limiter derived
by Levermore \& Pomraning (1981, eq. 28), which is given by
\begin{equation}
\lambda_{\rm FLD} = \frac{2+\mc{R}}{6+3\mc{R}+\mc{R}^2}\,.
\end{equation}
Its use in hydrodynamic simulations of star formation has been
documented, for example, in \citet{byrt90}, \citet{byl93,byl95}, and \citet{wb06}.

\subsection{Dust Opacities}\label{subsection:dust}

\begin{figure}
\epsscale{0.8} \plottwo{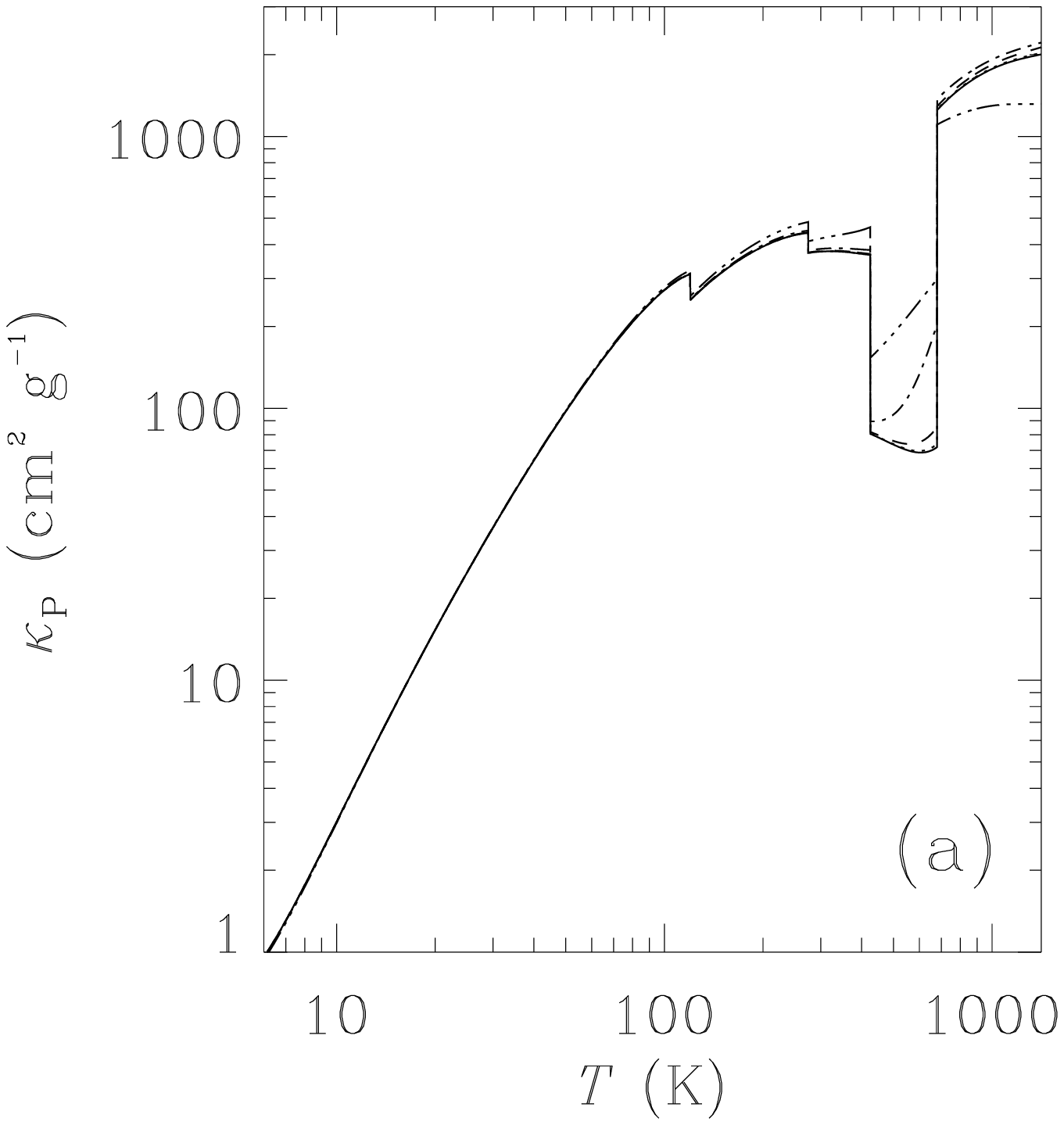}{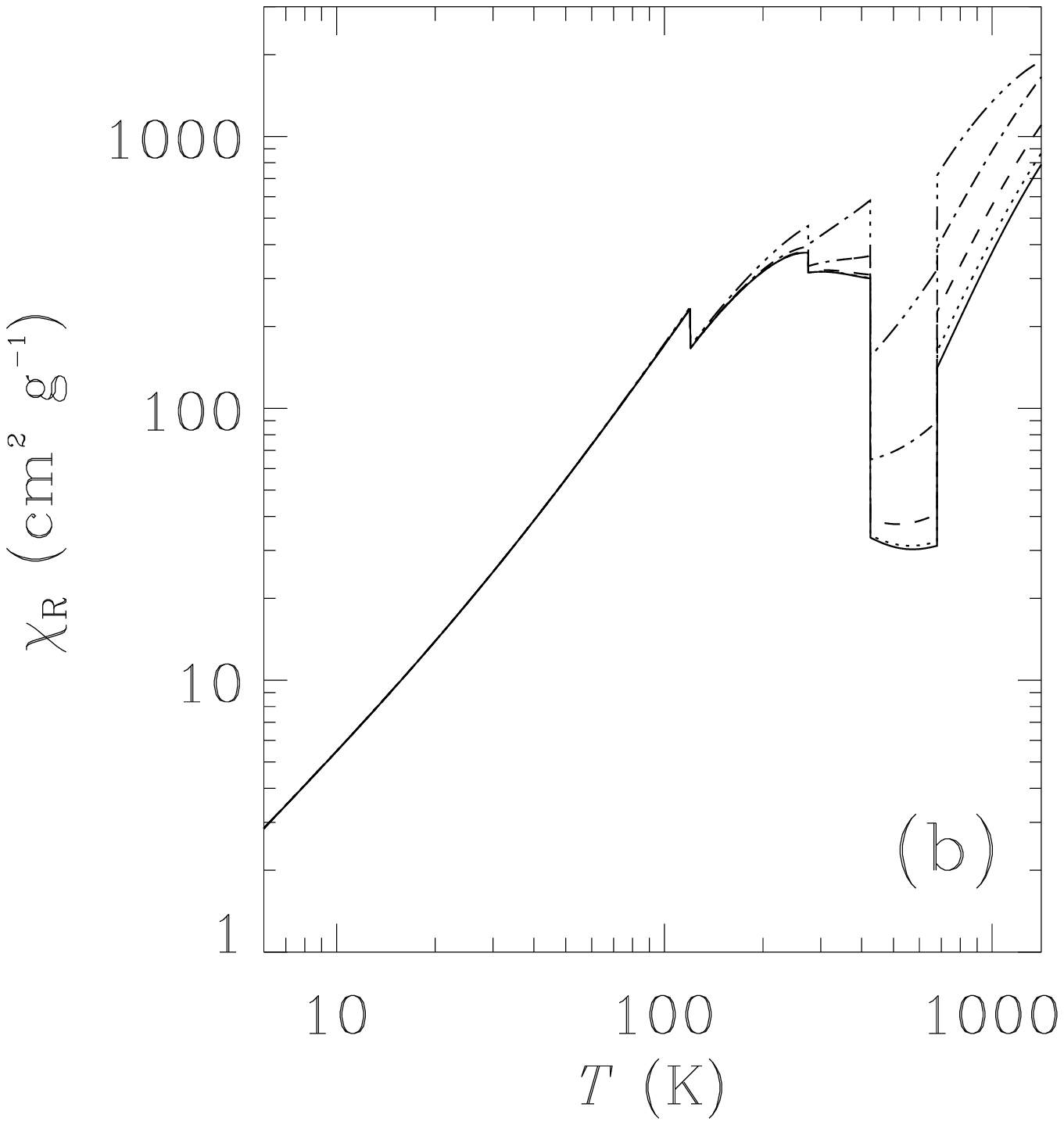}
\caption{(a) Planck mean absorption coefficients and (b) Rosseland
mean extinction coefficients (in cm$^2$ per gram of dust) for
grain sizes $a=0.0256\,\mu$m ({\em solid line}), $0.0543\,\mu$m
({\em dotted line}), $0.1190\,\mu$m ({\em dashed line}),
$0.2600\,\mu$m ({\em dash-dot line}), and $0.5680\,\mu$m ({\em
dash-dot-dot-dot line}).}\label{fig:opacity}
\end{figure}

For temperature less than $\simeq 1500$ K, the contribution of
dust to the total opacity dominates that from all other sources.
We take $\kappae = \kappap$ and $\chif = \chir$ (see Mihalas
\& Mihalas 1984, \S~82), where $\kappap$ and $\chir$ have been
obtained from private communication with Dmitry Semenov and Thomas
Henning. The major dust constituents are ``iron-poor" silicates,
troilite, organics, and water. Their relative mass fractions are
taken from \citet{phbsrf94}. These opacities (in cm$^2$ per gram
of dust) are shown in Figure \ref{fig:opacity} for the five
different grain size bins taken to represent an MRN distribution
(see \S~4.2 above). The major changes in the dust opacities are:
for temperatures $T<120$ K, all dust material are present; at
$T\simeq 120$ K, water ice evaporates; at $T=275$ K, volatile
organics evaporate; at $T=450$ K, refractory organics evaporate;
at $T=680$ K, troilite (FeS) evaporates.

\section{Modified Zeus-MP Code}\label{section:zeusmp}

In order to solve for the evolution of the many complex,
nonlinear systems of equations presented in this paper, numerical
techniques are necessary. Rather than extend the sophisticated
fully implicit, nonorthogonal adaptive mesh, two-fluid MHD code of
\citet{fm92} to include radiative transfer, we have opted instead
to modify the publicly-available Zeus-MP RMHD code
\citep{hayes06}. We have altered the algorithms governing the
evolution of the magnetic field in order to account for ambipolar
diffusion and Ohmic dissipation. In addition, we have added
routines to evolve the total grain density and to compute the
species abundances from the equilibrium chemical model detailed in
\S~\ref{subsection:chem}, and have made changes to Zeus-MP's
adaptive mesh module in order to track the collapsing core. New
modules were also written to improve both the efficiency of
Zeus-MP's implicit radiative transfer solver and the manner in
which the gravitational potential is calculated. A brief
description of these modifications follows.

The evolution of the magnetic field is governed by equation
(\ref{eqn:faraday2}). This equation does not assume flux-freezing
in any species and accounts for both ambipolar diffusion and Ohmic
dissipation. The first term on the right-hand side,
$\del\btimes(\vv{n}\btimes\bb{B})$, represents the advection of
the magnetic field by the neutrals. As long as the nonideal MHD
Courant condition
\begin{equation}
\Delta t \leq \Delta t_{\rm diff} \equiv
\frac{4\pi}{c^2\eta_\perp}\frac{(\Delta x)^2}{2}
\end{equation}
is satisfied, then the method of characteristics used to update
the magnetic field due to this term remains valid \citep{mlnkw95}
and we may use it without modification. Physically, this is
because, for a sufficiently short timestep, ambipolar diffusion
does not have time to alter the characteristics of Alfv\'{e}n
waves. The second term on the right-hand side,
$\del\btimes[(c^2\eta_\perp/4\pi)\del\btimes\bb{B}]$, represents
the diffusion of the magnetic field lines and is applied using the
method of constrained transport. This is done in the source step
part of the code. We employ a similar approach to the grain
continuity equation (\ref{eqn:gtcont3}). The right-hand side is
treated as a source term in the source step part of the code.
Then, the grain mass density is advected during the transport step
using the multi-species advection module already present in
Zeus-MP. Joule heating (eq. [\ref{eqn:joule}]) is also applied to
the internal energy during the source step.

We use an adaptive grid to track the evolution of the contracting
core. The grid, which must resolve the core, has its innermost
zone constrained so as to always have a width in the range
$\lambda_{\rm T,cr}/5 - \lambda_{\rm T,cr}/10$, where
$\lambda_{\rm T,cr} \equiv 1.4 C \tau_{\rm ff}$ is the critical
thermal lengthscale \citep{mouschovias91}, $C\equiv(k_{\rm
B}T/m_{\rm n})^{1/2}$ is the isothermal sound speed of the gas,
and $\tau_{\rm ff}\equiv(3\pi/32G\dr{n,c})^{1/2}$ is the spherical
free-fall time. (The quantity $\dr{n,c}$ is the neutral mass
density at the cloud center). The critical thermal lengthscale is
the smallest scale on which there can be spatial structure in the
density without thermal-pressure forces smoothing it out. The
number of cells is fixed, and their positions are spaced
logarithmically, so that the spacing between the $i$th and $i{\rm
th} + 1$ cells is a number greater than the spacing between the
$i{\rm th}-1$ and $i$th cells.

As discussed in \S~3.8.1 of \citet{hayes06}, Zeus-MP uses the
diagonal preconditioned conjugate gradient method to solve the
sparse matrix equation that results from spatially discretizing
equations (\ref{eqn:egcont}) and (\ref{eqn:ercont}). Diagonal
preconditioning is an attractive technique due to its simple
calculation, the fact that it poses no barrier to parallel
implementation, and its fairly common occurrence in linear
systems. However, it is only efficient for matrices in which the
main diagonal elements are much greater in magnitude than the
off-diagonal elements (a condition referred to as ``diagonal
dominance"). Unfortunately, this is generally not the case for the
problem studied here. We have therefore replaced the diagonal
preconditioner with an incomplete Cholesky decomposition
preconditioner, similar to what was provided in the public-release
version of Zeus-2D. The savings in computational cost has been
enormous.

Zeus-MP computes the gravitational potential through a two-step
process: first, the gravitational potential $\psi$ is found on the
computational boundaries; then $\psi$ is found in the interior by
iteratively solving the Poisson equation for $\psi$ using a sparse
matrix solver that relies on the preconditioned conjugate gradient
method. For each boundary of the domain, there are two possible
boundary types: (1) Neumann, in which the slope of the
gravitational potential is set to zero; and, (2) Dirichlet, in
which the value of $\psi$ in the ghost zones is specified. (There
is actually a third possible boundary type --- periodic boundary
conditions --- however, this boundary condition is not used here.)
Neumann boundary conditions are used at symmetry boundaries (axis
$r=0$ and equatorial plane $z=0$), while Dirichlet conditions are
applied at the outer boundaries, far from most of the mass
distribution. In the public-release version of Zeus-MP, Dirichlet
boundary conditions are implemented by computing $\psi$ on the
domain boundaries using a multipole expansion formula, which we
give here in spherical coordinates
$(\mathcalligra{r}\,,\theta,\phi)$ for an axisymmetric mass
distribution:
\begin{equation}\label{eqn:oldpsi}
\psi(\mathcalligra{r}\,,\theta) =
-G\sum_{\ell=0,2}\left[\int\rho(\mathcalligra{r}\,',\theta')P_\ell(\cos\theta')\mathcalligra{r}\,'^\ell
d^3
\mathcalligra{r}\,'\right]\frac{P_\ell(\cos\theta)}{\mathcalligra{r}\,^{\ell+1}}\,,
\end{equation}
where $P_\ell$ is the Legendre polynomial of degree $\ell$. Note
that Zeus-MP uses only the monopole and quadrupole moments, in
contrast to the earlier Zeus-2D code \citep{sn92}, which used
arbitrarily high $\ell$ moments until a desired convergence was
achieved. The term in brackets is denoted by $q_\ell$, and is
known as the multipole moment of order $\ell$ of the density
distribution. In most situations, a dozen or so multipole moments
are sufficient for convergence. They can be calculated once and
then used to find the potential at many boundary points. However,
we have found this subroutine inadequate for our purposes, as it
fails to converge in situations when the distance
$\mathcalligra{r}$ to the point at which one wishes to calculate
the potential is greater than the distance to any mass element. In
the axisymmetric geometry used in this work, this situation is
inevitable: mass elements near $(r,z) = (0,Z)$ or $(r,z)=(R,0)$
are closer to the origin than are mass elements near
$(r,z)=(R,Z)$. We therefore have followed \citet{dm01} in using
the more general multipole expansion \citep{jackson99}:
\begin{equation}\label{eqn:newpsi}
\psi(\mathcalligra{r}\,,\theta) =
-G\sum_{\ell=0}^\infty\left[\int\rho(\mathcalligra{r}\,',\theta')P_\ell(\cos\theta')\frac{\mathcalligra{r}\,^\ell_<}{\mathcalligra{r}\,^{\ell+1}_>}
d^3 \mathcalligra{r}\,'\right]P_\ell(\cos\theta)\,,
\end{equation}
where $\mathcalligra{r}\,_>$ ($\mathcalligra{r}\,_<$) is the
greater (lesser) of $\mathcalligra{r}\,'$ and
$\mathcalligra{r}\,$, and $\ell$ may take arbitrarily large
integral values until a desired convergence is achieved. An
unfortunate consequence of this more general expansion is that it
is {\em not} possible to perform one integration over all space
and use the result of that integration (the multipole moment
$q_\ell$) to find the potential at all boundary points. Instead, a
new integration over space must be performed for each boundary
point, since the location of that boundary point will determine
how the integral in equation (\ref{eqn:newpsi}) is separated into
two integrals: one integral will have $\mathcalligra{r}\,'$ in the
numerator of the integrand, and in the other integral
$\mathcalligra{r}\,'$ will be in the denominator. This situation
is further complicated by the use of parallelization, since
comparisons of $\mathcalligra{r}\,$ and $\mathcalligra{r}\,'$ and
subsequent integrations must take place across multiple
processors.

\section{Summary}\label{section:summary}

We have formulated the problem of the formation and evolution of
fragments (or cores) in magnetically-supported, self-gravitating
molecular clouds in two spatial dimensions. The evolution is
governed by the six-fluid RMHD equations. The magnetic flux is not
assumed to be frozen in any of the charged species. Its evolution
is determined by a newly-derived generalized Ohm's law, which
accounts for the contributions of both elastic and inelastic
collisions to ambipolar diffusion and Ohmic dissipation. The
species abundances (electrons, atomic and molecular ions,
positively-charged grains, negatively-charged grains, and neutral
grains) are calculated using an extensive equilibrium chemical
network. Both MRN and uniform grain size distributions are
considered. The thermal evolution of the protostellar core and its
effect on the dynamics are followed by employing the grey FLD
approximation. Realistic temperature-dependent grain opacities are
used that account for a variety of grain compositions. We have
augmented the publicly-available Zeus-MP code to take into
consideration all these effects and have modified several of its
algorithms to increase its accuracy and efficiency.

We summarize here for convenience the simplified evolutionary
equations discussed above and used in our modified version of the
Zeus-MP code:
\begin{subequations}
\begin{equation}
\D{t}{\dr{n}} + \del\bcdot(\dr{n}\vv{n}) = 0\,,
\end{equation}
\begin{equation}
\D{t}{\dr{g}} + \del\bcdot(\dr{g}\vv{n}) =
-\del\bcdot\bigl(\eta_{\rm cont,H}\bb{j}\btimes\eb\bigr)\,,
\end{equation}
\begin{equation}
\D{t}{(\dr{n}\vv{n})} + \del\bcdot(\dr{n}\vv{n}\vv{n}) = -\del
P_{\rm n} -\rho\del\psi +
\frac{1}{4\pi}(\del\btimes\bb{B})\btimes\bb{B} - \lambda_{\rm
FLD}\del\mc{E}\,,
\end{equation}
\begin{equation}
\D{t}{\bb{B}} = \del\btimes\left(\vv{n}\btimes\bb{B} -
\frac{c^2\eta_\perp}{4\pi}\del\btimes\bb{B}\right)\,,
\end{equation}
\begin{equation}
\del^2\psi = 4\pi G\dr{n}\,,
\end{equation}
\begin{equation}
\D{t}{\inteng} + \del\bcdot(\inteng\vv{n}) = -P_{\rm
n}\del\bcdot\vv{n} - 4\pi\kappap\mc{B} + c\kappap\mc{E}\,,
\end{equation}
\begin{equation}
\D{t}{\mc{E}} + \del\bcdot(\mc{E}\vv{n}) =
\del\bcdot\left(\frac{c\lambda_{\rm FLD}}{\chir}\del\mc{E}\right)
- \del\vv{n}\,\bb{:}\,\msb{P} + 4\pi\kappap\mc{B} -
c\kappap\mc{E}\,.
\end{equation}
\end{subequations}
These equations are considered together with the relations $P_{\rm
n}=(\gamma-1)\inteng$ and $\msb{P}=\msb{f}\mc{E}$. Results will be
presented in a forthcoming paper (Kunz \& Mouschovias 2009, in
preparation).

\section*{}
We thank Konstantinos Tassis, Vasiliki Pavlidou, Steve Desch,
Leslie Looney, and Duncan Christie for valuable discussions; John
Hayes for his assistance with the Zeus-MP code; and, Dmitry
Semenov and Thomas Henning for generously providing the dust
opacities. TM acknowledges partial support from the National
Science Foundation under grant NSF AST-07-09206.

\begin{appendix}

\section{Rate Coefficients}\label{app:ratecoeffs}

For radiative recombination of atomic ions and electrons,
$\alpha_{\rm rr}=2.8\times 10^{-12}\,(300\,{\rm K}/T)^{0.86}$
cm$^3$ s$^{-1}$; for the dissociative recombination of electrons
and HCO$^+$ ions, $\alpha_{\rm dr} = 2.0\times 10^{-7}\,(300\,{\rm
K}/T)^{0.75}$ cm$^3$ s$^{-1}$ \citep{un90}. The rate coefficient
adopted for charge exchange reactions between atomic and molecular
ions is $\beta = 2.5\times 10^{-9}$ cm$^3$ s$^{-1}$
\citep{watson76}.

The rate coefficients involving gas-phase species and grains are
taken from \citet{spitzer41,spitzer48}, with refinements made by
\citet{ds87} to account for the polarization of grains:
\begin{equation}
\alpha_{\rm eg_0} = \pi a^2\left(\frac{8k_{\rm B}T}{\pi m_{\rm
e}}\right)^{1/2}\left[1+\left(\frac{\pi e^2}{2ak_{\rm
B}T}\right)^{1/2}\right]\,\mc{P}_{\rm e}\,,
\end{equation}
\begin{equation}
\alpha_{\rm ig_0} = \pi a^2\left(\frac{8k_{\rm B}T}{\pi m_{\rm
i}}\right)^{1/2}\left[1+\left(\frac{\pi e^2}{2ak_{\rm
B}T}\right)^{1/2}\right]\,\mc{P}_{\rm i}\,,
\end{equation}
\begin{equation}
\alpha_{\rm eg_+} = \pi a^2\left(\frac{8k_{\rm B}T}{\pi m_{\rm
e}}\right)^{1/2}\left[1+\left(\frac{e^2}{ak_{\rm
B}T}\right)\right]\left[1+\left(\frac{2}{2+(ak_{\rm
B}T/e^2)}\right)^{1/2}\right]\,\mc{P}_{\rm e}\,,
\end{equation}
\begin{equation}
\alpha_{\rm ig_-} = \pi a^2\left(\frac{8k_{\rm B}T}{\pi m_{\rm
i}}\right)^{1/2}\left[1+\left(\frac{e^2}{ak_{\rm
B}T}\right)\right]\left[1+\left(\frac{2}{2+(ak_{\rm
B}T/e^2)}\right)^{1/2}\right]\,\mc{P}_{\rm i}\,.
\end{equation}
The sticking probabilities of electrons or ions onto grains,
denoted $\mc{P}_{\rm e}$ and $\mc{P}_{\rm i}$, are assigned the values
0.6 and 1.0, respectively \citep{umebayashi83}. Other quantities
in these equations have their usual meanings.

The rate coefficients for charge transfer between charged grains
are given by
\begin{equation}
\alpha_{{\rm g}^\alpha_+{\rm g}^{\alpha'}_-} = \pi a^2_{\rm
sum}\left(\frac{8k_{\rm B}T}{\pi m_{\rm
red}}\right)^{1/2}\left[1+\left(\frac{e^2}{a_{\rm sum}k_{\rm
B}T}\right)\right]\left[1+\left(\frac{2}{2+(a_{\rm sum}k_{\rm
B}T/e^2)}\right)^{1/2}\right]\,,
\end{equation}
\begin{equation}
\alpha_{{\rm g}^\alpha_\pm{\rm g}^{\alpha'}_0} = \pi a^2_{\rm
sum}\left(\frac{8k_{\rm B}T}{\pi m_{\rm
red}}\right)^{1/2}\left[1+\left(\frac{\pi e^2}{2a_{\rm sum}k_{\rm
B}T}\right)^{1/2}\right]\,\mc{P}_{\alpha\alpha'}\,,
\end{equation}
where the reduced mass of the two grains (labeled $\alpha$ and
$\alpha'$) is defined by $m_{\rm red} = m_\alpha
m_{\alpha'}/(m_\alpha + m_{\alpha'})$, and $a_{\rm
sum}=a_\alpha+a_{\alpha'}$ is the sum of the radii of two grains
$\alpha$ and $\alpha'$. The probability of two oppositely charged
grains neutralizing each other upon contact is assumed to be
unity. The probability of charge being transferred to a neutral
grain g$^{\alpha}_0$ from a charged grain g$^{\alpha'}_\pm$ is
assumed to be proportional to the surface areas of the grains, so
that $\mc{P}_{\alpha\alpha'} = a^2_{\alpha\alpha'}/(a^2_\alpha +
a^2_{\alpha'})$. In other words, of all the collisions between a
neutral grain, $\alpha$, and a charged grain, $\alpha'$, only a
fraction $\mc{P}_{\alpha}$ lead to charge exchange. The
complementary probability $\mc{P}_{\alpha'}$ leaves the charges
unchanged.

\section{Generalized Ohm's Law}\label{app:gol}

\subsection{Derivation}\label{app:ohmslaw}

We consider equations (\ref{eqn:sforce2}) and
(\ref{eqn:gzforce3}), repeated here for convenience, which are to
be solved for the species drift velocities $\bb{w}_s$ relative to
the neutrals:
\begin{equation}\label{eqn:ohmeqn1}
0 =
\frac{\omega_s\tausn{s}}{1+\varrho_s}\left(\frac{c}{B}\bb{E}_{\rm
n}+\ws{s}\btimes\eb\right) - \ws{s} +
\frac{\varrho_s}{1+\varrho_s}\ww{g_0}\,,
\end{equation}
\begin{equation}\label{eqn:ohmeqn2}
0 = \ww{g_0} - \sum_k\frac{\tau_0}{\tsinel{k}}\ws{k}\,.
\end{equation}
We define, for brevity and clarity of presentation, the following
quantities
\begin{subequations}
\begin{equation}
\Psi_{1,s} = {{\displaystyle
\frac{\tau_0}{\tsinel{s}}\frac{\omega_s\tausn{s}}{1+\varrho_s}}\over{\displaystyle\denomone}}\,,\qquad
\Psi_1 = \sum_k \Psi_{1,k}\,;
\end{equation}
\begin{equation}
\Psi_{2,s} = {{\displaystyle
\frac{\tau_0}{\tsinel{s}}\frac{\omega^2_s\tausn{s}^2}{(1+\varrho_s)^2}}\over{\displaystyle\denomone}}\,,\qquad
\Psi_2 = \sum_k \Psi_{2,k}\,;
\end{equation}
\begin{equation}
\Psi_{3,s} = {{\displaystyle
\frac{\tau_0}{\tsinel{s}}\frac{\omega_s\tausn{s}}{1+\varrho_s}\frac{\varrho_s}{1+\varrho_s}}\over{\displaystyle\denomone}}\,,\qquad
\Psi_3 = \sum_k \Psi_{3,k}\,.
\end{equation}
\end{subequations}
We recall that the index $k$ runs over all the charged species
independently of the index $s$, which denotes the charged species
in question. Note that the denominator in the above expressions
may be written with the help of equation (\ref{eqn:tauz}b) as
\begin{equation}
\frac{\tau_0}{\taurn{g_0}} + \sum_k
\frac{\tau_0}{\tsinel{k}}\frac{1}{1+\varrho_k}\,,
\end{equation}
which shows its positive definite nature.

We first multiply equation (\ref{eqn:ohmeqn1}) by
$\tau_0/\tsinel{s}$, sum over $s$, and use equation
(\ref{eqn:ohmeqn2}) to find that
\begin{equation}\label{eqn:ohmeqn3}
\ww{g_0} = \Psi_1\frac{c}{B}\bb{E}_{\rm n} +
\sum_k\Psi_{1,k}\ws{k}\btimes\eb\,,
\end{equation}
where we have switched the summation index to $k$ to avoid
confusion with the species in question, $s$. Next we take the
cross product of equations (\ref{eqn:ohmeqn1}) and
(\ref{eqn:ohmeqn3}) and the unit vector $\eb$:
\begin{equation}\label{eqn:ohmeqn4}
\ws{s}\btimes\eb =
\frac{\omega_s\tausn{s}}{1+\varrho_s}\left(\frac{c}{B}\bb{E}_{\rm
n}\btimes\eb - \ws{s,\perp}\right) +
\frac{\varrho_s}{1+\varrho_s}\ww{g_0}\btimes\eb\,,
\end{equation}
\begin{equation}\label{eqn:ohmeqn5}
\ww{g_0}\btimes\eb = \Psi_1\frac{c}{B}\bb{E}_{\rm n}\btimes\eb -
\sum_k\Psi_{1,k}\ws{k,\perp}\,.
\end{equation}
Equation (\ref{eqn:ohmeqn5}) is now substituted in equation
(\ref{eqn:ohmeqn4}) to obtain
\begin{equation}\label{eqn:ohmeqn6}
\ws{s}\btimes\eb =
\left(\frac{\omega_s\tausn{s}}{1+\varrho_s}+\frac{\varrho_s}{1+\varrho_s}\Psi_1\right)\frac{c}{B}\bb{E}_{\rm
n}\btimes\eb -
\left(\frac{\omega_s\tausn{s}}{1+\varrho_s}\ws{s,\perp}+\frac{\varrho_s}{1+\varrho_s}\sum_k\Psi_{1,k}\ws{k,\perp}\right)\,.
\end{equation}
Inserting this expression in equation (\ref{eqn:ohmeqn3}), we find
that
\begin{equation}\label{eqn:ohmeqn7}
\ww{g_0} = \Psi_1\frac{c}{B}\bb{E}_{\rm n} + \bigl(\Psi_2 +
\Psi_3\Psi_1\bigr)\frac{c}{B}\bb{E}_{\rm n}\btimes\eb -
\sum_k\bigl(\Psi_{2,k}+\Psi_3\Psi_{1,k}\bigr)\ws{k,\perp}\,,
\end{equation}
which is now ready to be inserted, along with equation
(\ref{eqn:ohmeqn6}), in equation (\ref{eqn:ohmeqn1}):
\begin{align}\label{eqn:ohmeqn}
\ws{s} &+
\sum_k\left[\frac{\omega_s\tausn{s}}{1+\varrho_s}\left(\frac{\omega_k\tausn{k}}{1+\varrho_k}\delta_{sk}
+ \frac{\varrho_s}{1+\varrho_s}\Psi_{1,k}\right) +
\frac{\varrho_s}{1+\varrho_s}(\Psi_{2,k}+\Psi_3\Psi_{1,k})\right]\ws{k,\perp}\nonumber\\*
&=\left(\frac{\omega_s\tausn{s}}{1+\varrho_s}+\frac{\varrho_s}{1+\varrho_s}\Psi_1\right)\frac{c}{B}\bb{E}_{\rm
n} +
\left[\frac{\omega_s\tausn{s}}{1+\varrho_s}\left(\frac{\omega_s\tausn{s}}{1+\varrho_s}+\frac{\varrho_s}{1+\varrho_s}\Psi_1\right)
+
\frac{\varrho_s}{1+\varrho_s}\bigl(\Psi_2+\Psi_3\Psi_1\bigr)\right]\frac{c}{B}\bb{E}_{\rm
n}\btimes\eb\,.
\end{align}
The symbol $\delta_{sk}$ is the Kronecker delta. This is our first
main result: it gives the velocity of each charged species in
terms of the electric field in the frame of the neutrals. Another
way of interpreting this equation is obtained by defining the
velocity of the magnetic field lines with respect to the lab
frame:
\begin{equation}\label{eqn:vf}
\vv{f}\equiv\frac{c}{B}\bb{E}\btimes\eb\,.
\end{equation}
Then equation (\ref{eqn:ohmeqn}) provides the velocities of all
the charged species in terms of the neutral velocity and the
field-line velocity. We made use of this concept earlier in
Section \ref{subsection:attachment}.

Equation (\ref{eqn:ohmeqn}) can be separated into components
parallel and perpendicular to the magnetic field. The parallel
component of the current density is easily obtained:
\begin{subequations}\label{eqn:jparl}
\begin{align}
\bb{j}_{||} &= \sum_s n_s q_s \ws{s,||} \\
            &= \sum_s n_s q_s \left(\frac{\omega_s\tausn{s}}{1+\varrho_s}+\frac{\varrho_s}{1+\varrho_s}\Psi_1\right)\frac{c}{B}\bb{E}_{\rm n,||} \\
            &= \sum_s \sigma_s(1-\varsigma_s)\bb{E}_{\rm n,||} \\
            &\equiv \sigma_{||}\bb{E}_{\rm n,||}\,,
\end{align}
\end{subequations}
where we have introduced the conductivity of species $s$,
$\sigma_s = n_s q^2_s \tausn{s}/m_s$, and $\varsigma_s$, given in
Appendix \ref{app:defs}, is the factor by which the conductivity
of species $s$ is altered because of inelastic collisions. In the
last step above, we have introduced the parallel conductivity,
$\sigma_{||}$, which is defined {\em in situ}. Note that
$\varsigma_s\ge 0$ for all $s$. In other words, by interfering
with the rate at which the charge carriers flow along the magnetic
field, inelastic collisions are responsible for decreasing
(increasing) the parallel conductivity (resistivity) of the gas.

Finding the perpendicular components of the current density is not
as straightforward and amounts to solving a matrix equation. We
first define the $4\times 1$ column vectors $\bb{C}^\perp$ and
$\bb{C}^{\rm H}$,  whose entries are given by
\begin{equation}
\renewcommand{\theequation}{\ref{app:gol}\arabic{equation}a,b}
\bigl(\bb{C}^\perp\bigr)_s =
\omega_s\tausn{s}(1-\varsigma_s)\qquad{\rm and}\qquad
\bigl(\bb{C}^{\rm H}\bigr)_s =
-\omega^2_s\tausn{s}^2(1-\varpi_s)\,.
\end{equation}
We also define the $4\times 4$ matrix of coefficients $\msb{A}$
whose entries are given by
\begin{equation}\label{eqn:Amatrix}
\bigl(\msb{A}\bigr)_{sk} =
\bigl[1+\omega^2_s\tausn{s}^2(1-\varphi_s)\bigr]\delta_{sk} +
\omega^2_k\tausn{k}^2\vartheta_{sk}(1-\delta_{sk})\,.
\end{equation}
The expressions for $\varsigma_s$, $\varpi_s$, $\varphi_s$, and
$\vartheta_{sk}$ are given below in Appendix \ref{app:defs}. Then
the perpendicular component of equation (\ref{eqn:ohmeqn}) takes
on the form
\begin{equation}\label{eqn:ohmmatrix}
\bb{C}^\perp \frac{c}{B}\bb{E}_{\rm n,\perp} - \bb{C}^{\rm H}
\frac{c}{B}\bb{E}_{\rm n}\btimes\eb = \msb{A}\bb{W}_\perp\,,
\end{equation}
where $\bb{W}_\perp$ is the $4\times 1$ column vector of unknown
velocities of charge species relative to neutrals, $[\ww{e}$,
$\ww{i}$, $\ww{g_-}$, $\ww{g_+}]^\intercal$.

We use Cramer's method to solve the matrix equation
(\ref{eqn:ohmmatrix}). We define
\begin{equation}
D = \det\bigl[\msb{A}\bigr]\,.
\end{equation}
In addition, we use the notation $D^\perp_s$ to represent the
determinant of $\msb{A}$ with the $s$th column of $\msb{A}$ having
been replaced by $\bb{C}^\perp$. Similarly, $D^{\rm H}_s$ is the
determinant of $\msb{A}$ with the $s$th column having been
replaced by $\bb{C}^{\rm H}$. Then, the solution of the system
(\ref{eqn:ohmmatrix}) is
\begin{equation}
\ws{s,\perp} = \frac{D^\perp_s}{D}\frac{c}{B}\bb{E}_{\rm n,\perp}
- \frac{D^{\rm H}_s}{D}\frac{c}{B}\bb{E}_{\rm n}\btimes\eb\,.
\end{equation}
Once the determinants have been computed, the current density
perpendicular to the magnetic field may be obtained:
\begin{subequations}\label{eqn:jperp}
\begin{align}
\bb{j}_\perp &= \sum_s n_s q_s \ws{s,\perp} \\
             &= \frac{\sum_s n_s q_s D^\perp_s}{D}\frac{c}{B}\bb{E}_{\rm n,\perp} - \frac{\sum_s n_s q_s D^{\rm H}_s}{D}\frac{c}{B}\bb{E}_{\rm n}\btimes\eb \\
             &= \sum_s \frac{\sigma_s(1-\varsigma_s)}{1+\omega^2_s\tausn{s}^2(1-\varphi_s)}\Upsilon(\varsigma)\bb{E}_{\rm n,\perp} + \sum_s \frac{\sigma_s\omega_s\tausn{s}(1-\varpi_s)}{1+\omega^2_s\tausn{s}^2(1-\varphi_s)}\Upsilon(\varpi)
             \bb{E}_{\rm n}\btimes\eb \\
             &\equiv \sigma_\perp\bb{E}_{\rm n,\perp} - \sigma_{\rm H}\bb{E}_{\rm n}\btimes\eb\,.
\end{align}
\end{subequations}
In the last step, we have defined the perpendicular conductivity
$\sigma_\perp$ and the Hall conductivity $\sigma_{\rm H}$, which
include the effects of inelastic collisions.

\subsection{Modification due to a Grain Size Distribution}\label{app:grainsize}

When considering a grain size distribution, rather than
single-size grains, two changes must be made to the above
derivation. First, the inelastic collision timescales given by
equations (\ref{eqn:tgpinel}) and (\ref{eqn:tgminel}) must be
modified as follows:
\begin{subequations}
\begin{equation}\label{eqn:tgpinel2}
\trinel{g_+} \rightarrow \trinel{g^{\it \alpha}_+} =
\left[\frac{1}{\trinel{g^{\it \alpha}_0i}} + \frac{\dr{g^{\it
\alpha}_+}}{\dr{g^{\it \alpha}_0}}\frac{1}{\trinel{g^{\it
\alpha}_+e}} + \sum_{\alpha'}\left(\frac{1}{\trinel{g^{\it
\alpha}_0g^{\it \alpha'}_+}} + \frac{\dr{g^{\it
\alpha}_+}}{\dr{g^{\it \alpha}_0}}\frac{1}{\trinel{g^{\it
\alpha}_+g^{\it \alpha'}_0}} + \frac{\dr{g^{\it
\alpha}_+}}{\dr{g^{\it \alpha}_0}}\frac{1}{\trinel{g^{\it
\alpha}_+g^{\it \alpha'}_-}}\right)\right]^{-1}\,,
\end{equation}
\begin{equation}\label{eqn:tgminel2}
\trinel{g_-} \rightarrow \trinel{g^{\it \alpha}_-} =
\left[\frac{1}{\trinel{g^{\it \alpha}_0e}} + \frac{\dr{g^{\it
\alpha}_-}}{\dr{g^{\it \alpha}_0}}\frac{1}{\trinel{g^{\it
\alpha}_-i}} + \sum_{\alpha'}\left(\frac{1}{\trinel{g^{\it
\alpha}_0g^{\it \alpha'}_-}} + \frac{\dr{g^{\it
\alpha}_-}}{\dr{g^{\it \alpha}_0}}\frac{1}{\trinel{g^{\it
\alpha}_-g^{\it \alpha'}_0}} + \frac{\dr{g^{\it
\alpha}_-}}{\dr{g^{\it \alpha}_0}}\frac{1}{\trinel{g^{\it
\alpha}_-g^{\it \alpha'}_+}}\right)\right]^{-1}\,.
\end{equation}
\end{subequations}
The summations over $\alpha'$ (the grain size label) indicate that
a grain of size $\alpha$ may give or receive charges not only from
other grains of its own size, but also from all other different-size grains.
Second, the summation index $s$ in equations (\ref{eqn:jparl}) and
(\ref{eqn:jperp}) should range over {\em all} charged species,
including all sizes of charged grains.

\section{Derivation of Species Velocities}\label{app:velocities}

En route to the derivation of a generalized Ohm's law, the
differential velocity of every species can be obtained in terms of
the current density:
\begin{align}\label{eqn:velocitiesj}
n_s q_s \ws{s} &= \sigma_{||,s}\bb{E}_{\rm n,||} +
\sigma_{\perp,s}\bb{E}_{\rm n,\perp}
               - \sigma_{{\rm H},s}\bb{E}_{\rm n}\btimes\eb\nonumber \\
               &= \sigma_{||,s}\eta_{||}\bb{j}_{||} + \sigma_{\perp,s}\bigl(\eta_\perp\bb{j}_\perp + \eta_{\rm H}\bb{j}\btimes\eb\bigr) - \sigma_{{\rm H},s}\bigl(\eta_\perp\bb{j}\btimes\eb - \eta_{\rm H}\bb{j}_\perp\bigr)\nonumber \\
               &= \sigma_{||,s}\eta_{||}\bb{j}_{||} + \bigl(\sigma_{\perp,s}\eta_\perp + \sigma_{{\rm H},s}\eta_{\rm H}\bigr)\bb{j}_\perp + \bigl(\sigma_{\perp,s}\eta_{\rm H} - \sigma_{{\rm H},s}\eta_\perp\bigr)\bb{j}\btimes\eb\,.
\end{align}
Using equation (\ref{eqn:vf}), is is straightforward to show that
\begin{equation}
\ww{f} \equiv \vv{f} - \vv{n} =
\frac{c\eta_\perp}{B}\bb{j}\btimes\eb - \frac{c\eta_{\rm
H}}{B}\bb{j}_\perp\,.
\end{equation}
We may then write the components of the current density in terms
of the differential velocity of the field lines as
\begin{equation}
\renewcommand{\theequation}{\ref{app:velocities}\arabic{equation}a,b}
\frac{c}{B}\bb{j}\btimes\eb = \sigma_\perp\ww{f,\perp} -
\sigma_{\rm H}\ww{f}\btimes\eb\qquad{\rm and}\qquad
-\frac{c}{B}\bb{j}_\perp = \sigma_{\rm H}\ww{f,\perp} +
\sigma_\perp\ww{f}\btimes\eb\,.
\end{equation}
Defining the indirect coupling coefficient $\Theta_s$ implicitly
by
\begin{equation}
\frac{\Theta_s}{\Theta_s+1} \equiv \left(\frac{B}{cn_s
q_s}\right)\bigl[\sigma_\perp(\sigma_{\perp,s}\eta_{\rm H} -
\sigma_{{\rm H},s}\eta_\perp) - \sigma_{\rm
H}(\sigma_{\perp,s}\eta_\perp + \sigma_{{\rm H},s}\eta_{\rm
H})\bigr]\,,
\end{equation}
and introducing
\begin{equation}
\Lambda_s \equiv - \left(\frac{B}{cn_s
q_s}\right)\bigl[\sigma_\perp(\sigma_{\perp,s}\eta_\perp +
\sigma_{{\rm H},s}\eta_{\rm H}) + \sigma_{\rm
H}(\sigma_{\perp,s}\eta_{\rm H} - \sigma_{{\rm
H},s}\eta_\perp)\bigr]\,,
\end{equation}
equation (\ref{eqn:velocitiesj}) may now be written in component
form as
\begin{equation}
\renewcommand{\theequation}{\ref{app:velocities}\arabic{equation}a,b}
    \ws{s,\perp} = \frac{\Theta_s}{\Theta_s+1}\ww{f,\perp} + \Lambda_s \ww{f}\btimes\eb\qquad{\rm and}\qquad
\ws{s}\btimes\eb = \frac{\Theta_s}{\Theta_s+1}\ww{f}\btimes\eb -
\Lambda_s\ww{f,\perp}\,,
\end{equation}
or, more explicitly,
\begin{subequations}
\begin{align}
\vs{s,\perp} &= \vv{n,\perp} \frac{1}{\Theta_s+1} + \vv{f,\perp} \frac{\Theta_s}{\Theta_s+1} +(\vv{f}-\vv{n})\btimes\eb\,\Lambda_s\,, \\
\vs{s}\btimes\eb &= \vv{n}\btimes\eb \frac{1}{\Theta_s+1} +
\vv{f}\btimes\eb \frac{\Theta_s}{\Theta_s+1} -
(\vv{f,\perp}-\vv{n,\perp})\,\Lambda_s\,.
\end{align}
\end{subequations}
These equations were discussed in Section
\ref{subsection:attachment}.

\section{Definitions}\label{app:defs}

In the main text, as well as in the preceding appendices, we had
delayed giving explicit definitions of $\varsigma_s$, $\varpi_s$,
$\varphi_s$, $\Upsilon_s$, and $\vartheta_{sk}$ for all $(s,k)$ =
e, i, g$_-$, and g$_+$ due to their complexity and length. Here we
give explicit expressions for these quantities for all the charged
species. Before we proceed, however, a few simplifications are in
order. Since both $(\trinel{e}/\taurn{e})$ and
$(\trinel{i}/\taurn{i})\gg 1$ for the density regime of interest
in this paper, we may neglect the influence of inelastic
collisions on the electron and ion fluids. Using the results of
\citet{tm07b}, we also may assume that the velocity difference
between a given grain and a neutral particle is less than the
sound speed of the gas. These are both excellent assumptions, and
lead to a much more compact form of the following definitions than
would otherwise be possible.

The variable $\varsigma_s$ first appeared in the definition of the
parallel conductivity (\ref{eqn:jparl}) and again later in the
definition of the perpendicular conductivity (\ref{eqn:jperp}).
For electrons and ions, $\varsigma_{\rm e}=\varsigma_{\rm i} = 0$,
because of the negligible influence of inelastic collisions on the
electron and ion fluids relative to that of elastic collisions.
The expressions for the negative and positive grains are given by
\begin{equation}
\varsigma_{\rm g_\pm} = \frac{\varrho_{\rm g_\pm}}{1+\varrho_{\rm
g_\pm}} \left[{{\displaystyle
\frac{\tau_0}{\taurn{g_0}}+\frac{\tau_0}{\trinel{g_\mp}}\frac{2}{1+\varrho_{\rm
g\mp}}}\over{\displaystyle \denomtwo}}\right]\,,
\end{equation}
and are clearly positive. As mentioned in Section
\ref{subsection:ohmslaw}, by interfering with the rate at which
the charge carriers flow along the magnetic field, inelastic
collisions are responsible for decreasing (increasing) the
parallel conductivity (resistivity) of the gas.

The derivation of the perpendicular conductivity involved many
more definitions, all of which are given below. For the same
reason stated above for which $\varsigma_{\rm e} = \varsigma_{\rm
i} = 0$, the expressions for $\varpi_s$ and $\varphi_s$ vanish
when $s$ = e or i. The quantity $\Upsilon_s$ is equal to unity for
these species. The nontrivial $\varpi_s$, $\varphi_s$, and
$\Upsilon_s$ for $s$ = g$_-$, g$_+$ are given by
\begin{equation}
\varpi_{\rm g_\pm} = \frac{\varsigma_{\rm g_\pm}}{1+\varrho_{\rm
g_\pm}} + \frac{\varrho_{\rm g_\mp}}{1+\varrho_{\rm g_\mp}}
\left[{{\displaystyle \frac{\tau_0}{\taurn{g_0}} +
\frac{\tau_0}{\trinel{g_+}}\frac{\varsigma_{\rm
g_+}}{1+\varrho_{\rm g_+}} +
\frac{\tau_0}{\trinel{g_-}}\frac{\varsigma_{\rm
g_-}}{1+\varrho_{\rm
g_-}}}\over{\displaystyle\denomtwo}}\right]\,;
\end{equation}
\begin{align}
\varphi_{\rm g_\pm} &= \frac{\varrho_{\rm g_\pm}}{1+\varrho_{\rm
g_\pm}}\frac{2+\varrho_{\rm g_\pm}}{1+\varrho_{\rm g_\pm}}
\left[{{\displaystyle \frac{\tau_0}{\taurn{g_0}}+\frac{\tau_0}{\trinel{g_\mp}}\frac{1}{1+\varrho_{\rm g\mp}}}\over{\displaystyle \denomtwo}}\right]\nonumber \\
&+ \frac{\tau_0}{\trinel{g_\pm}}\frac{\varrho_{\rm
g_\pm}}{(1+\varrho_{\rm g_\pm})^2} \left[{{\displaystyle
\frac{\tau_0}{\taurn{g_0}}\frac{\varrho_{\rm
g_\pm}}{1+\varrho_{\rm
g_\pm}}+\frac{\tau_0}{\trinel{g_\mp}}\frac{1}{1+\varrho_{\rm
g\mp}}\left(\frac{\varrho_{\rm g_+}}{1+\varrho_{\rm g_+}} +
\frac{\varrho_{\rm g_-}}{1+\varrho_{\rm
g_-}}\right)}\over{\displaystyle
\left(\denomtwo\right)^2}}\right]\,;
\end{align}
\begin{equation}
\Upsilon_{\rm g_\pm}(\varsigma) = {{\displaystyle 1 +
\frac{\omega^2_{\rm g_\mp}\tausn{g_\mp}^2\vartheta_{\rm g_\pm
g_\mp}}{1+\omega^2_{\rm g_\mp}\tausn{g_\mp}^2(1-\varphi_{\rm
g_\mp})}\frac{1-\varsigma_{\rm g_\mp}}{1-\varsigma_{\rm
g_\pm}}}\over{\displaystyle \denomthree}}\,;
\end{equation}
\begin{equation}
\Upsilon_{\rm g_\pm}(\varpi) = {{\displaystyle 1 +
\frac{\omega^2_{\rm g_\mp}\tausn{g_\mp}^2\vartheta_{\rm g_\pm
g_\mp}}{1+\omega^2_{\rm g_\mp}\tausn{g_\mp}^2(1-\varphi_{\rm
g_\mp})}\frac{1-\varpi_{\rm g_\mp}}{1-\varpi_{\rm
g_\pm}}}\over{\displaystyle \denomthree}}\,.
\end{equation}
In equation (\ref{eqn:Amatrix}), we had introduced
$\vartheta_{sk}$ as a measure of the inelastic collisional
coupling between different pairs of charged species, $s\ne k$.
This variable was also used in the definition of $\Upsilon_s$
above. Since the effect of inelastic collisions on the electron
and ion fluids is negligible, $\vartheta_{sk}$ vanishes for
$(s,k)$ = e or i. The only remaining nonzero values of
$\vartheta_{sk}$ involve the charged grain species, and are given
by
\begin{equation}
\vartheta_{\rm g_\pm g_\mp} = \frac{\varrho_{\rm
g_\pm}}{1+\varrho_{\rm
g_\pm}}\frac{\tau_0}{\trinel{g_\mp}}\frac{1}{1+\varrho_{\rm
g_\mp}}\left[{{\displaystyle \frac{1}{1+\varrho_{\rm
g_\mp}}\left(1-\frac{\tau_0}{\trinel{g_\pm}}\frac{\varrho_{\rm
g_\pm}}{1+\varrho_{\rm g_\pm}}\right) - \frac{1}{1+\varrho_{\rm
g_\pm}}\left(1-\frac{\tau_0}{\trinel{g_\mp}}\frac{\varrho_{\rm
g_\mp}}{1+\varrho_{\rm g_\mp}}\right)}\over{\displaystyle
\left(\denomtwo\right)^2}}\right]\,.
\end{equation}

\end{appendix}


\begin{thebibliography}{99}

\bibitem[Adams \& Shu(1985)]{as85}
Adams, F., \& Shu, F. H. 1985, \apj, 296, 655

\bibitem[Adams \& Shu(1986)]{as86}
Adams, F., \& Shu, F. H. 1986, \apj, 308, 836

\bibitem[Alves et al.(2008)]{afg08}
Alves, F. O., Franco, G. A. P., \& Girart, J. M. 2008, A\&A, 486,
L13

\bibitem[Arons \& Max(1975)]{am75}
Arons, J., \& Max, C. E. 1975, \apj, 196, L77

\bibitem[Babcock \& Cowling(1953)]{bc53}
Babcock, H. W., \& Cowling, T. G. 1953, \mnras, 113, 357

\bibitem[Bacmann et al.(2000)]{bapabwt00}
Bacmann, A., Andr\'{e}, P., Puget, J. L., Abergel, A., Bontemps,
S., \& Ward-Thompson, D. 2000, A\&A, 361, 555

\bibitem[Baker(1979)]{baker79}
Baker, P. L. 1979, A\&A, 75, 54

\bibitem[Basu \& Mouschovias(1994)]{bm94}
Basu, S., \& Mouschovias, T. Ch. 1994, \apj, 432, 720

\bibitem[Basu \& Mouschovias(1995a)]{bm95a}
Basu, S., \& Mouschovias, T. Ch. 1995a, \apj, 452, 386

\bibitem[Basu \& Mouschovias(1995b)]{bm95b}
Basu, S., \& Mouschovias, T. Ch. 1995b, \apj, 453, 271

\bibitem[Bate(1998)]{bate98}
Bate, M. R. 1998, \apj, 508, L95

\bibitem[Baudry et al.(1981)]{bpddc81}
Baudry, A., Perault, M., de La Noe, J., Despois, D., Cernicharo,
J. 1981, A\&A, 104, 101

\bibitem[Black \& Bodenheimer(1975)]{bb75}
Black, D. C., \& Bodenheimer, P. 1975, \apj, 199, 619

\bibitem[Black \& Bodenheimer(1976)]{bb76}
Black, D. C., \& Bodenheimer, P. 1976, \apj, 206, 138

\bibitem[Bodenheimer(1968)]{bodenheimer68}
Bodenheimer, P. 1968, \apj, 153, 483

\bibitem[Bodenheimer et al.(1993)]{byl93}
Bodenheimer, P., Yorke, H. W., \& Laughlin, G. 1993, \apj, 411,
274

\bibitem[Bodenheimer et al.(1995)]{byl95}
Bodenheimer, P., Yorke, H. W., \& Laughlin, G. 1995, \apj, 443,
199

\bibitem[Bodenheimer et al.(1990)]{byrt90}
Bodenheimer, P., Yorke, H. W., R\'{o}\.{z}yszka, M., \& Tohline,
J. E. 1990, \apj, 355, 651

\bibitem[Boss(1981)]{boss81}
Boss, A. P. 1981, \apj, 250, 636

\bibitem[Boss(1984)]{boss84}
Boss, A. P. 1984, \apj, 277, 768

\bibitem[Boss(1986)]{boss86}
Boss, A. P. 1986, \apjs, 62, 519

\bibitem[Boss(1988)]{boss88}
Boss, A. P. 1988, \apj, 331, 370

\bibitem[Boss \& Myhill(1995)]{bm95}
Boss, A. P., \& Myhill, E. A. 1995, 451, 218

\bibitem[Boss \& Yorke(1990)]{by90}
Boss, A. P., \& Yorke, H. W. 1990, \apj, 353, 236

\bibitem[Chandrasekhar \& Fermi(1953)]{cf53}
Chandrasekhar, S., \& Fermi, E. 1953, \apj, 118, 116

\bibitem[Ciolek \& K\"{o}nigl(1998)]{ck98}
Ciolek, G. E., \& K\"{o}nigl, A. 1998, \apj, 504, 257

\bibitem[Ciolek \& Mouschovias(1993)]{cm93}
Ciolek, G. E., \& Mouschovias, T. Ch. 1993, \apj, 418, 774

\bibitem[Ciolek \& Mouschovias(1994)]{cm94}
Ciolek, G. E., \& Mouschovias, T. Ch. 1994, \apj, 425, 142

\bibitem[Ciolek \& Mouschovias(1995)]{cm95}
Ciolek, G. E., \& Mouschovias, T. Ch. 1995, \apj, 454, 194

\bibitem[Ciolek \& Mouschovias(1996)]{cm96}
Ciolek, G. E., \& Mouschovias, T. Ch. 1996, \apj, 468, 749

\bibitem[Ciolek \& Mouschovias(1998)]{cm98}
Ciolek, G. E., \& Mouschovias, T. Ch. 1998, \apj, 504, 280

\bibitem[Clayton \& Leising(1987)]{cl87}
Clayton, D. D., \& Leising, M. D. 1987, Phys. Rep., 144, 1

\bibitem[Consolmagno \& Jokipii(1978)]{cj78}
Consolmagno, G. J., \& Jokipii, J. R. 1978, The Moon and the
Planets, 19, 253

\bibitem[Cortes \& Crutcher(2006)]{cc06}
Cortes, P. C., \& Crutcher, R. M. 2006, \apj, 639, 965

\bibitem[Cortes et al.(2005)]{ccw05}
Cortes, P. C., Crutcher, R. M., \& Watson, W. D. 2005, \apj, 628,
780

\bibitem[Crutcher(1999)]{crutcher99}
Crutcher, R. M. 1999, \apj, 520, 706

\bibitem[Crutcher \& Kaz\`{e}s(1983)]{ck83}
Crutcher, R. M., \& Kaz\`{e}s, I. 1983, A\&A, 125, L23

\bibitem[Crutcher et al.(1994)]{cmtc94}
Crutcher, R. M., Mouschovias, T. Ch., Troland, T. H., \& Ciolek,
G. E. 1994, \apj, 427, 839

\bibitem[Crutcher et al.(2004)]{cnwtk04}
Crutcher, R. M., Nutter, D. J., Ward-Thompson, D., \& Kirk, J. M.
2004, \apj, 600, 279

\bibitem[Crutcher et al.(1996)]{crmt96}
Crutcher, R. M., Roberts, D. A., Mehringer, D. M., \& Troland, T.
H. 1996, \apj, 462, L79

\bibitem[Crutcher et al.(1999a)]{crtg99}
Crutcher, R. M., Roberts, D. A., Troland, T. H., \& Goss, W. M.
1999a, \apj, 515, 275

\bibitem[Crutcher et al.(1993)]{ctghkm93}
Crutcher, R. M., Troland, T. H., Goodman, A. A., Heiles, C.,
Kaz\`{e}s, I., \& Myers, P. C. 1993, \apj, 407, 175

\bibitem[Crutcher et al.(1987)]{ctk87}
Crutcher, R. M., Troland, T. H., \& Kaz\`{e}s, I. 1987, A\&A, 181,
119

\bibitem[Crutcher et al.(1999b)]{ctlpk99}
Crutcher, R. M., Troland, T. H., Lazareff, B., Paubert, G., \&
Kaz\`{e}s, I. 1999b, \apj, 514, L121

\bibitem[Desch \& Mouschovias(2001)]{dm01}
Desch, S. J., \& Mouschovias, T. Ch. 2001, \apj, 550, 314

\bibitem[Draine \& Sutin(1987)]{ds87}
Draine, B. T., \& Sutin, B. 1987, \apj, 320, 803

\bibitem[Elmegreen(1979)]{elmegreen79}
Elmegreen, B. G. 1979, \apj, 232, 729

\bibitem[Elmegreen(1986)]{elmegreen86}
Elmegreen, B. G. 1986, in Light on Dark Matter, ed. F. P. Israel
(Dordrecht: Reidel), 265

\bibitem[Elmegreen \& Scalo(2004)]{es04}
Elmegreen, B. G., \& Scalo, J. 2004, \araa, 42, 211

\bibitem[Eng(2002)]{eng02}
Eng, C. 2002, Ph.D. Thesis, University of Illinois at
Urbana-Champaign

\bibitem[Fielder \& Mouschovias(1992)]{fm92}
Fiedler, R. A., \& Mouschovias, T. Ch. 1992, \apj, 391, 199

\bibitem[Fielder \& Mouschovias(1993)]{fm93}
Fiedler, R. A., \& Mouschovias, T. Ch. 1993, \apj, 415, 680

\bibitem[Gaustad(1963)]{gaustad63}
Gaustad, J. E. 1963, \apj, 138, 1050

\bibitem[Girart et al.(1999)]{gcr99}
Girart, J. M., Crutcher, R. M., \& Rao, R. 1999, \apj, 525, L109

\bibitem[Girart et al.(2006)]{grm06}
Girart, J. M., Rao, R., \& Marrone, D. P. 2006, Science, 313, 812

\bibitem[Glassgold(1995)]{glassgold95}
Glassgold, A. E. 1995, \apj, 438, L111

\bibitem[Glassgold \& Langer(1974)]{gl74}
Glassgold, A. E., \& Langer, W. D. 1974, \apj, 193, 73

\bibitem[Goldreich \& Reisenegger(1992)]{gr92}
Goldreich, P., \& Reisenegger, A. 1992, \apj, 395, 250

\bibitem[Goldsmith \& Arquilla(1985)]{ga85}
Goldsmith, P. F., \& Arquilla, R. 1985, in Protostars and Planets
II, ed. D. C. Black \& M. S. Matthews (Tucson: University of
Arizona Press), 137

\bibitem[Goodman et al.(1989)]{gchmt89}
Goodman, A. A., Crutcher, R. M., Heiles, C., Myers, P. C., \&
Troland, T. H. 1989, \apj, 338, L61

\bibitem[Hayashi(1966)]{hayashi66}
Hayashi, C. 1966, \araa, 4, 171

\bibitem[Hayes et al.(2006)]{hayes06}
Hayes, J. C., Norman, M. L., Fiedler, R. A., Bordner, J. O., Li,
P. S., Clark, S. E., ud-Doula, A., \& Mac Low, M.-M. 2006, \apjs,
165, 188

\bibitem[Heiles(1987)]{heiles87}
Heiles, C. 1987, in Physical Processes in Interstellar Clouds, ed.
G. E. Morfill \& M. Scholer (Dordrecht: Reidel), 429

\bibitem[Heiles \& Crutcher(2005)]{hc05}
Heiles, C., \& Crutcher, R. 2005, in Cosmic Magnetic Fields, ed.
R. Wielebinski \& R. Beck (Berlin: Springer), 137

\bibitem[Heyer et al.(1987)]{hvsssgs87}
Heyer, M. H., Vrba, F. J., Snell, R. L., Schloerb, F. P., Strom,
S. E., Goldsmith, P. F., \& Strom, K. M. 1987, \apj, 321, 855

\bibitem[Hildebrand et al.(1999)]{hddsv99}
Hildebrand, R. H., Dotson, J. L., Dowell, C. D., Schleuning, D.
A., \& Vaillancourt, J. E. 1999, \apj, 516, 834

\bibitem[Hollenbach et al.(1971)]{hws71}
Hollenbach, D. J., Werner, M. W., \& Salpeter, E. E. 1971, \apj,
163, 165

\bibitem[Jackson(1999)]{jackson99}
Jackson, J. D. 1999, Classical Electrodynamics (3rd ed.; New York:
John Wiley \& Sons)

\bibitem[Kaz\`{e}s \& Crutcher(1986)]{kc86}
Kaz\`{e}s, I., \& Crutcher, R. M. 1986, A\&A, 164, 328

\bibitem[Kittel(1958)]{kittel58}
Kittel, C. 1958, Elementary Statistical Physics (New York: John
Wiley \& Son)

\bibitem[Lai et al.(2001)]{lcgr01}
Lai, S.-P., Crutcher, R. M., Girart, J. M., \& Rao, R. 2001, \apj,
561, 864

\bibitem[Lai et al.(2002)]{lcgr02}
Lai, S.-P., Crutcher, R. M., Girart, J. M., \& Rao, R. 2002, \apj,
566, 925

\bibitem[Lai et al.(2003)]{lgc03}
Lai, S.-P., Girart, J. M., \& Crutcher, R. M. 2003, \apj, 598, 392

\bibitem[Larson(1969)]{larson69}
Larson, R. B. 1969, \mnras, 145, 271

\bibitem[Larson(1972)]{larson72}
Larson, R. B. 1972, \mnras, 157, 121

\bibitem[Larson(1981)]{larson81}
Larson, R. B. 1981, \mnras, 194, 809

\bibitem[Lequeux(1975)]{lequeux75}
Lequeux, J. 1975, A\&A, 39, 257

\bibitem[Levermore \& Pomraning(1981)]{lp81}
Levermore, C. D., \& Pomraning, G. C. 1981, \apj, 248, 321

\bibitem[McKee(1989)]{mckee89}
McKee, C. F. 1989, \apj, 345, 782

\bibitem[Mac Low \& Klessen(2004)]{mlk04}
Mac Low, M.-M., \& Klessen, R. S. 2004, Rev. Mod. Phys., 76, 125

\bibitem[Mac Low et al.(1995)]{mlnkw95}
Mac Low, M.-M., Norman, M. L., K\"{o}nigl, A., \& Wardle, M. 1995,
\apj, 442, 726

\bibitem[Masunaga et al.(1998)]{mmi98}
Masunaga, H., Miyama, S. M., \& Inutsuka, S.-i. 1998, \apj, 495,
346

\bibitem[Mathis et al.(1977)]{mrn77}
Mathis, J. S., Rumpl, W., \& Nordsieck, K. H. 1977, \apj, 217, 425

\bibitem[Mellon \& Li(2008)]{ml08}
Mellon, R. R., \& Li, Z.-Y. 2008, astro-ph/0809.3593

\bibitem[Mestel(1965)]{mestel65}
Mestel, L. 1965, QJRAS, 6, 265

\bibitem[Mestel(1966)]{mestel66}
Mestel, L. 1966, \mnras, 133, 265

\bibitem[Mestel \& Spitzer(1956)]{ms56}
Mestel, L., \& Spitzer, L., Jr. 1956, \mnras, 116, 503

\bibitem[Mihalas \& Mihalas(1984)]{mm84}
Mihalas, D., \& Mihalas, B. W. 1984, Foundations of Radiation
Hydrodynamics (New York: Oxford University Press)

\bibitem[Morton(1974)]{morton74}
Morton, D. C. 1974, \apj, 193, L35

\bibitem[Mouschovias(1976a)]{mouschovias76a}
Mouschovias, T. Ch. 1976a, \apj, 206, 753

\bibitem[Mouschovias(1976b)]{mouschovias76b}
Mouschovias, T. Ch. 1976b, \apj, 207, 141

\bibitem[Mouschovias(1977)]{mouschovias77}
Mouschovias, T. Ch. 1977, \apj, 211, 147

\bibitem[Mouschovias(1978)]{mouschovias78}
Mouschovias, T. Ch. 1978, in Protostars and Planets, ed. T.
Gehrels (Tucson: University of Arizona Press), 209

\bibitem[Mouschovias(1979)]{mouschovias79}
Mouschovias, T. Ch. 1979, \apj, 228, 159

\bibitem[Mouschovias(1987)]{mouschovias87}
Mouschovias, T. Ch. 1987, in Physical Processes in Interstellar
Clouds, ed. G. E. Morfill \& M. Scholer (Dordrecht: Reidel), 453

\bibitem[Mouschovias(1991)]{mouschovias91}
Mouschovias, T. Ch. 1991, \apj, 373, 169

\bibitem[Mouschovias(1996)]{mouschovias96}
Mouschovias, T. Ch. 1996, in Solar and Astrophysical
Magnetohydrodynamic Flows, ed. K. Tsiganos (Dordrecht: Kluwer),
505

\bibitem[Mouschovias \& Morton(1991)]{mm91}
Mouschovias, T. Ch., \& Morton, S. A. 1991, \apj, 371, 296

\bibitem[Mouschovias \& Morton(1992a)]{mm92a}
Mouschovias, T. Ch., \& Morton, S. A. 1992a, \apj, 390, 144

\bibitem[Mouschovias \& Morton(1992b)]{mm92b}
Mouschovias, T. Ch., \& Morton, S. A. 1992b, \apj, 390, 166

\bibitem[Mouschovias \& Paleologou(1979)]{mp79}
Mouschovias, T. Ch., \& Paleologou, E. V. 1979, \apj, 230, 204

\bibitem[Mouschovias \& Paleologou(1980)]{mp80}
Mouschovias, T. Ch., \& Paleologou, E. V. 1980, \apj, 237, 877

\bibitem[Mouschovias et al.(1985)]{mpf85}
Mouschovias, T. Ch., Paleologou, E. V., \& Fiedler, R. A. 1985,
\apj, 291, 772

\bibitem[Mouschovias \& Psaltis(1995)]{mp95}
Mouschovias, T. Ch., \& Psaltis, D. 1995, \apj, 444, L105

\bibitem[Mouschovias \& Spitzer(1976)]{ms76}
Mouschovias, T. Ch., \& Spitzer, L., Jr. 1976, \apj, 210, 326

\bibitem[Mouschovias et al.(2006)]{mtk06}
Mouschovias, T. Ch., Tassis, K., \& Kunz, M. W. 2006, \apj, 646,
1043

\bibitem[Myers(1985)]{myers85}
Myers, P. C. 1985, in Protostars \& Planets II, ed. D. C. Black \&
M. S. Matthews (Tucson: University of Arizona Press), 81

\bibitem[Myers \& Benson(1983)]{mb83}
Myers, P. C., \& Benson, P. J. 1983, \apj, 266, 309

\bibitem[Myers \& Gammie(1999)]{mg99}
Myers, P. C., \& Gammie, C. F. 1999, \apj, 522, L141

\bibitem[Myers et al.(1983)]{mlb83}
Myers, P. C., Linke, R. A., \& Benson, P. J. 1983, \apj, 264, 517

\bibitem[Nakano(1979)]{nakano79}
Nakano, T. 1979, PASJ, 31, 697

\bibitem[Nakano \& Umebayashi(1980)]{nu80}
Nakano, T., \& Umebayashi, T. 1980, PASJ, 32, 613

\bibitem[Nakano \& Umebayashi(1986a)]{nu86a}
Nakano, T., \& Umebayashi, T. 1986a, \mnras, 218, 663

\bibitem[Nakano \& Umebayashi(1986b)]{nu86b}
Nakano, T., \& Umebayashi, T. 1986b, \mnras, 221, 319

\bibitem[Nishi et al.(1991)]{nnu91}
Nishi, R., Nakano, T., \& Umebayashi, T. 1991, \apj, 368, 181

\bibitem[Novak et al.(1997)]{nddgh97}
Novak, G., Dotson, J. L., Dowell, C. D., Goldsmith, P. F., \&
Hildebrand, R. H. 1997, \apj, 487, 320

\bibitem[Novak et al.(1989)]{nghpd89}
Novak, G., Gonatas, D. P., Hildebrand, R. H., Platt, S. R., \&
Dragovan, M. 1989, \apj, 345, 802

\bibitem[Paleologou \& Mouschovias(1983)]{pm83}
Paleologou, E. V., \& Mouschovias, T. Ch. 1983, \apj, 275, 838

\bibitem[Parks(1991)]{parks91}
Parks, G. K. 1991, Physics of Space Plasmas (Redwood City:
Addison-Wesley), 283

\bibitem[Pneuman \& Mitchell(1965)]{pm65}
Pneuman, G. W., \& Mitchell, T. P. 1965, Icarus, 4, 494

\bibitem[Pollack et al.(1994)]{phbsrf94}
Pollack, J. B., Hollenbach, D., Beckwith, S., Simonelli, D. P.,
Roush, T., \& Fong, W. 1994, \apj, 421, 615

\bibitem[Preibisch et al.(1995)]{psy95}
Preibisch, Sonnhalter, \& Yorke, H. W. 1995, A\&A, 299, 144

\bibitem[Schleuning(1998)]{schleuning98}
Schleuning, D. A. 1998, \apj, 493, 811

\bibitem[Schleuning et al.(2000)]{svhdndd00}
Schleuning, D. A., Vaillancourt, J. E., Hildebrand, R. H., Dowell,
C. D., Novak, G., Dotson, J. L., \& Davidson, J. A. 2000, \apj,
535, 913

\bibitem[Scott \& Black(1980)]{sb80}
Scott, E. H., \& Black, D. C. 1980, ApJ, 239, 166

\bibitem[Snow(1976)]{snow76}
Snow, T. P., Jr. 1976, \apj, 204, 759

\bibitem[Spitzer(1941)]{spitzer41}
Spitzer, L., Jr. 1941, \apj, 93, 369

\bibitem[Spitzer(1948)]{spitzer48}
Spitzer, L., Jr. 1948, \apj, 107, 6

\bibitem[Spitzer(1968)]{spitzer68}
Spitzer, L., Jr. 1968, Diffuse Matter in Space (New York:
Interscience)

\bibitem[Spitzer(1978)]{spitzer78}
Spitzer, L., Jr. 1978, Physical Processes in the Interstellar
Medium (New York: Wiley-Interscience)

\bibitem[Stone et al.(1992)]{smn92}
Stone, J. M., Mihalas, D., \& Norman, M. L. 1992, \apjs, 80, 819

\bibitem[Stone \& Norman(1992)]{sn92}
Stone, J. M., \& Norman, M. L. 1992, \apjs, 80, 753

\bibitem[Strittmatter(1966a)]{strittmatter66a}
Strittmatter, P. A. 1966a, \mnras, 131, 491

\bibitem[Strittmatter(1966b)]{strittmatter66b}
Strittmatter, P. A. 1966b, \mnras, 132, 359

\bibitem[Tassis \& Mouschovias(2005a)]{tm05a}
Tassis, K., \& Mouschovias, T. Ch. 2005a, \apj, 618, 769

\bibitem[Tassis \& Mouschovias(2005b)]{tm05b}
Tassis, K., \& Mouschovias, T. Ch. 2005b, \apj, 618, 783

\bibitem[Tassis \& Mouschovias(2007a)]{tm07a}
Tassis, K., \& Mouschovias, T. Ch. 2007a, \apj, 660, 370

\bibitem[Tassis \& Mouschovias(2007b)]{tm07b}
Tassis, K., \& Mouschovias, T. Ch. 2007b, \apj, 660, 388

\bibitem[Tassis \& Mouschovias(2007c)]{tm07c}
Tassis, K., \& Mouschovias, T. Ch. 2007c, \apj, 660, 402

\bibitem[Troland et al.(1986)]{tck86}
Troland, T. H., Crutcher, R. M., \& Kaz\`{e}s, I. 1986, \apj, 304,
L57

\bibitem[Troland et al.(1996)]{tcghkm96}
Troland, T. H., Crutcher, R. M., Goodman, A. A., Heiles, C.,
Kaz\`{e}s, I., \& Myers, P. C. 1996, \apj, 471, 302

\bibitem[Tscharnuter(1975)]{tscharnuter75}
Tscharnuter, W. 1975, A\&A, 39, 207

\bibitem[Tscharnuter(1978)]{tscharnuter78}
Tscharnuter, W. 1978, The Moon and the Planets, 19, 229

\bibitem[Tscharnuter \& Winkler(1979)]{tw79}
Tscharnuter, W. M., \& Winkler, K.-H. A. 1979, Comput. Phys.
Comm., 18, 171

\bibitem[Turner \& Stone(2001)]{ts01}
Turner, N. J., \& Stone, J. M. 2001, \apjs, 135, 95

\bibitem[Umebayashi(1983)]{umebayashi83}
Umebayashi, T. 1983, Prog. Theor. Phys., 69, 480

\bibitem[Umebayashi \& Nakano(1980)]{un80}
Umebayashi, T., \& Nakano, T. 1980, PASJ, 32, 405

\bibitem[Umebayashi \& Nakano(1990)]{un90}
Umebayashi, T., \& Nakano, T. 1990, \mnras, 243, 103

\bibitem[Vrba et al.(1981)]{vct81}
Vrba, F. J., Coyne, G. V., \& Tapia, S. 1981, \apj, 243, 489

\bibitem[Watson(1976)]{watson76}
Watson, W. D. 1976, Rev. Mod. Phys., 48, 513

\bibitem[Whitehouse \& Bate(2006)]{wb06}
Whitehouse, S. C., \& Bate, M. R. 2006, \mnras, 367, 32

\bibitem[Winkler \& Newman(1980a)]{wn80a}
Winkler, K.-H., \& Newman, M. J. 1980a, \apj, 236, 201

\bibitem[Winkler \& Newman(1980b)]{wn80b}
Winkler, K.-H., \& Newman, M. J. 1980b, \apj, 238, 311

\bibitem[Yorke(1977)]{yorke77}
Yorke, H. W. 1977, A\&A, 58, 423

\bibitem[Yorke(1980)]{yorke80}
Yorke, H. W. 1980, A\&A, 85, 215

\bibitem[Yorke \& Krugel(1977)]{yk77}
Yorke, H. W., \& Krugel, E. 1977, A\&A, 54, 183

\bibitem[Yorke \& Shustov(1981)]{ys81}
Yorke, H. W., \& Shustov, B. M. 1981, A\&A, 98, 125

\bibitem[Yorke \& Sonnhalter(2002)]{ys02}
Yorke, H. W., \& Sonnhalter, C. 2002, \apj, 569, 846

\bibitem[Zuckerman \& Evans(1974)]{ze74}
Zuckerman, B., \& Evans, N. J. 1974, \apj, 192, L149

\bibitem[Zweibel \& Josafatsson(1983)]{zj83}
Zweibel, E. G., \& Josafatsson, K. 1983, \apj, 270, 511

\end{thebibliography}
\end{document}